\documentclass[%
 aip, jcp,
 amsmath,amssymb,
 reprint,%
]{revtex4-1}

\usepackage{graphicx}
\usepackage{dcolumn}
\usepackage{bm}
\usepackage{longtable}
\usepackage[utf8]{inputenc}
\usepackage[T1]{fontenc}
\usepackage{mathptmx}
\usepackage{etoolbox}

\makeatletter
\def\@email#1#2{%
 \endgroup
 \patchcmd{\titleblock@produce}
  {\frontmatter@RRAPformat}
  {\frontmatter@RRAPformat{\produce@RRAP{*#1\href{mailto:#2}{#2}}}\frontmatter@RRAPformat}
  {}{}
}%
\makeatother
\begin{document}

\preprint{AIP/123-QED}

\title[]{On phase separation and crystallization of Ge-rich GeSbTe alloys from atomistic simulations with a machine learning interatomic potential}

\author{Omar Abou El Kheir}
\author{Dario Baratella}
\author{Marco Bernasconi}%
 \email{marco-bernasconi@unimib.it}
\affiliation{ 
Department of Materials Science, University of Milano-Bicocca, Via R. Cozzi 55, I-20125 Milano, Italy
}%

\date{\today}

\begin{abstract}
We developed a machine learning interatomic potential  (MLIP) for Ge-rich GeSbTe alloys of interest for applications in  phase change memories embedded in microcontrollers. The MLIP was generated by fitting with a neural network method a large database of energies and forces computed within density functional theory of elemental, binary, stoichiometric and non-stoichiometric ternary alloys in the Ge-Sb-Te  phase diagram. The MLIP is demonstrated to be highly transferable to large regions of the phase diagram around the compositions included in the dataset. The MLIP is then exploited to simulate the crystallization with  phase separation of three Ge-rich alloys on the Ge-Sb$_2$Te$_3$ and Ge- Ge$_2$Sb$_2$Te$_5$ tie-lines that correspond to  the set process of the memory cell. The transformation on the ns time scale and at 600 K, comparable to the operation conditions of the memory,  yields crystalline cubic GeTe slightly Sb-doped  and amorphous GeSb and Ge.  These metastable phases differ from the thermodynamically stable products and form due to kinetics effects on the short time span of the set operation in phase change memories.
\end{abstract}

\maketitle

\section{\label{sec:intro} Introduction}
Phase change memories (PCMs) are of interest for several applications, including stand-alone storage class memories \cite{fantini,BoniardiRedaelli}, in-memory and neuromorphic computing  
devices \cite{sebastian2,seyedChemRev}, and embedded memories in microcontrollers \cite{Cappelletti2020,redaelli2022}.
These applications exploit a fast and reversible transformation between the crystalline and amorphous phases of a chalcogenide alloy, typically GeSbTe (GST)
\cite{wuttig,noe2017phase,zhang2019designing}, induced by heating. Depending on the device architecture,  
heat is provided by current pulses either via Joule effects within the transforming material or via a resistive element, the heater, in contact
 with a  thin film of GST. By applying an intense current pulse, crystalline GST is brought above the melting temperature and subsequently amorphizes due to fast cooling (reset process). By applying a less intense and longer current pulse, the amorphous phase recrystallizes (set process). 
 
 For embedded applications, the amorphous phase must sustain the thermal budget for the soldering process in which the device is exposed to 550 K for a few minutes. To meet this requirement, Ge-rich GST alloys (GGST) have been developed reaching a crystallization temperature T$_{\rm x}$ above 
  600 K for alloys on the Ge-Sb$_2$Te$_3$ tie-line 
\cite{ZULIANI1,ZULIANI201527,palumbo2017,navarro2013trade}, which is much higher than that (T$_{\rm x}$=420-440 K \cite{NOE2016}) of the flagship Ge$_2$Sb$_2$Te$_5$ (GST225) compound,  mostly used for neuromorphic and stand-alone devices \cite{wuttig,noe2017phase,zhang2019designing}.
GGST has already been exploited in the next-generation microcontrollers (28 nm MOS technology) \cite{arnaud2018,arnaud2020}.
 
 The increase in T$_{\rm x}$ comes at the cost of increased complexity in the set and reset processes.
 In fact, the crystallization of non-stoichiometric GGST alloys involves the segregation of Ge in excess and the formation of GST crystallites with a lower fraction of Ge \cite{ZULIANI201527,palumbo2017,agati,luong2021some,privitera2018atomic,privitera2020crystallization,Prazakova,fattorini,cecchi2022crystallization,Yimam2022}. 
 The mass transport that accompanies this transformation slows  the overall process, increasing T$_{\rm x}$.  The crystallization temperature can be further increased by doping with nitrogen \cite{cheng2012,NavarroNGGST2016,prazakova2020,remondina2025,luong2021some,THOMAS2021} which was shown to reduce the mobility of Ge \cite{Elliott2015}.
 
 For GGST alloys on the Ge-GST225 tie-line, which are often discussed in literature \cite{Prazakova,privitera2018atomic,privitera2020crystallization,rahier}, the transformation should yield pure Ge and GST225 from the thermodynamical point of view. In fact, in the ternary Ge-Sb-Te phase diagram, the only thermodynamically stable compounds lie on the GeTe-Sb$_2$Te$_3$ and Sb-Sb$_2$Te$_3$ tie-lines
 \cite{OmarHigh}. Indeed, heating GGST films for hours above T$_{\rm x}$ leads to the formation of face centered cubic (fcc) GST with a lattice parameter consistent with cubic GST on the GeTe-Sb$_2$Te$_3$ pseudobinary line \cite{agati}.
 
 However, in the short time span of the set operation in the devices, kinetic effect might favor the formation of metastable phases or intermediate products preventing the transformation into  the thermodynamically stable compounds. 
For example, experimental evidence  was provided  on the formation of GeTe as an intermediate product in the decomposition of GGST films\cite{privitera2020crystallization,rahier}. In Ref. \cite{rahier}, Ge segregation and  the formation of GeTe crystals were observed  annealing a GGST film above 310 $^o$C. First, GeTe crystallizes in the  metastable $Pnma$ phase that triggers the crystallization of pure Ge and then evolves into the cubic phase. Only by further annealing at 410 $^o$C, incorporation of Sb leads to the formation of crystallites with composition close to GST225 \cite{rahier}. In other works in which the temperature was probably sufficiently higher than T$_x$, the final products consisting of crystalline Ge and cubic GST were observed with Ge crystallizing first \cite{Prazakova}.

The likely possibility of formation of cubic GeTe in the decomposition process of GGST was actually predicted   by density functional theory (DFT) calculation of the convex hull  in the Ge-Sb-Te phase diagram \cite{OmarHigh}. The knowledge of the DFT formation energy of all GST alloys in the central part of the
Ge-Sb-Te ternary phase diagram allowed us to identify the metastable cubic crystalline phases, in addition to the themodynamically stable compounds on the GeTe-Sb$_2$Te$_3$ tie-line,  that are more
likely to form during the crystallization of a generic GST alloy \cite{OmarHigh}. It was shown that Sb-doped cubic GeTe was the most probable product in the decomposition of several Ge-rich alloys on the Ge-Sb$_2$Te$_3$, Ge-GST225 and Ge-GST124 tie-lines \cite{Omar2021,cecchi2022crystallization,Yimam2022}.

In the real device, the transformation is further affected by the presence of temperature gradients and high electric fields that drive atoms with effective charges of different sign in opposite directions. An example of the complex nanostructure arising
in the GGST-PCM is provided in Ref. \cite{petroni2025} where scanning transmission electron microscopy (TEM) and electron energy loss spectroscopy (EELS) were used to characterize a PCM cell after the forming operation, i.e. the first cycle of melting and recrystallization of an active dome embedded in a polycrystalline matrix, resulting in turn from the uniform heating of the as-deposited amorphous homogeneous alloy. This analysis revealed a complex nanostructure formed by a core more Sb-rich and a Ge-rich shell  with evidence of formation of GeTe and trigonal Sb crystallites \cite{petroni2025}.
However, many details on the role of the electric field, thermal gradient, and morphology of the  polycrystalline matrix embedding the transforming material are still unclear.

In this respect, atomistic simulations can provide useful information on the parameters that control possible different decomposition pathways in this complex system. 
We have already mentioned that DFT calculations provided information on the decomposition propensity of GGST alloys and the most likely decomposition path during crystallization  based on thermochemical data  \cite{OmarHigh}. The stability of homogeneous  GGST alloys with respect to phase separation in still amorphous products was also addressed by DFT molecular dynamics (MD) in Ref. \cite{sun2021ab} where it was shown that segregation of Ge in the amorphous phase is expected for a Ge content greater than 50 $\%$. However, these latter works only considered the thermodynamical aspects of the phase transformation and not possible kinetics effects due to the limitations in the size and time scale of DFT-MD simulations.
These limitations can be overcome by MD simulations with machine learning interatomic potentials (MLIP) generated by training over a large database (DB) of DFT energies \cite{Behler,gap1,Jacobs2025,ReviewMLIP2025}. 
MLIP simulations of the crystallization process have been reported in literature for some stoichiometric phase change alloys including GeTe \cite{SossoCryst,Sosso2015} and GST225 \cite{npjOmar,Omar2025,DeringerNatComm2025}.

In a previous work,  we have shown that crystallization and decomposition of simpler binary Ge-rich Ge$_x$Te alloys can be simulated on the ns time scale by MLIP-MD \cite{baratella2025}.
On the other hand, we have also utilized a MLIP for stoichiometric GST225 to simulate the set process in a multimillion atom model of a PCM \cite{Omar2025} in the Wall architecture typically used for embedded memories \cite{arnaud2020}.

In the perspective of simulating the set process of embedded memories at the device scale, here we develop a MLIP for GGST alloys by leveraging our previous works on Ge-rich Ge$_x$Te \cite{baratella2025} and stoichiometric GST225 \cite{npjOmar} and using the DeePMD code \cite{DeePMD4,DeePMD2}. The neural network (NN) scheme implemented in DeePMD was trained on a DFT-DB including the elemental, stoichiometric binary and ternary compounds, 
and the Ge-rich Ge$_5$Sb$_2$Te$_3$ (GST523) alloy on the Ge-Sb$_2$Te$_3$ tie-line. The training database also includes the DBs of the stoichiometric GST225 \cite{npjOmar} and the Ge-rich  Ge$_2$Te alloys \cite{baratella2025}, generated previously. The GST523 alloy was studied in previous experimental \cite{ZULIANI201527,Yimam2022} and theoretical \cite{Omar2021} works, and was also shown to be the constituent of the active region of PCMs cells in the reset state after forming (first melting and crystallization of the end-of-fabrication memory cell) from alloys richer in Ge (see Fig. 7 in Ref. \cite{navarro2020}). 

Since we aim at simulating the decomposition process of the alloy, the MLIP must be able to reproduce not only the starting and the final compositions, but also the intermediate phases/compositions that the system could visit during the transformation process.  The MLIP must then be transferable to compositions not present in the training DB. We will show that
the MLIP is indeed capable of describing alloys within a large region around GST523 in the ternary phase diagram. This includes
the triangular region
with  the GST523, GeTe and Ge$_2$Te alloys,  all  present in the training DB,  at the vertices (see Fig. \ref{fig:validation}a).

The MLIP was then utilized to simulate the crystallization with phase separation of GST523 on the Ge-Sb$_2$Te$_3$ tie-line, of the   Ge$_7$Sb$_2$Te$_5$ (GST725) alloy  with the same fraction of Ge (50 $\%$), but lying on the Ge-GST225 tie-line, and finally of the Ge$_{9.{\bar 6}}$Sb$_2$Te$_5$ alloy which lies very close to the intersection between the GST523-Ge$_2$Te  and the Ge-GST225 lines  (see Fig. \ref{fig:validation}a).
Although thermodynamics predicts that the products of the crystallization would consist of the extremes of the tie-lines
of the ternary phase diagram, kinetics effects lead to the formation of intermediate products that are the only phases visible on the ns time scale  which is comparable to, albeit shorter than, the time span of the set process in the device.

\section{Methods}\label{sec:methods}
\begin{figure*}[]
    \centering
    \includegraphics[width=0.8\linewidth,keepaspectratio]{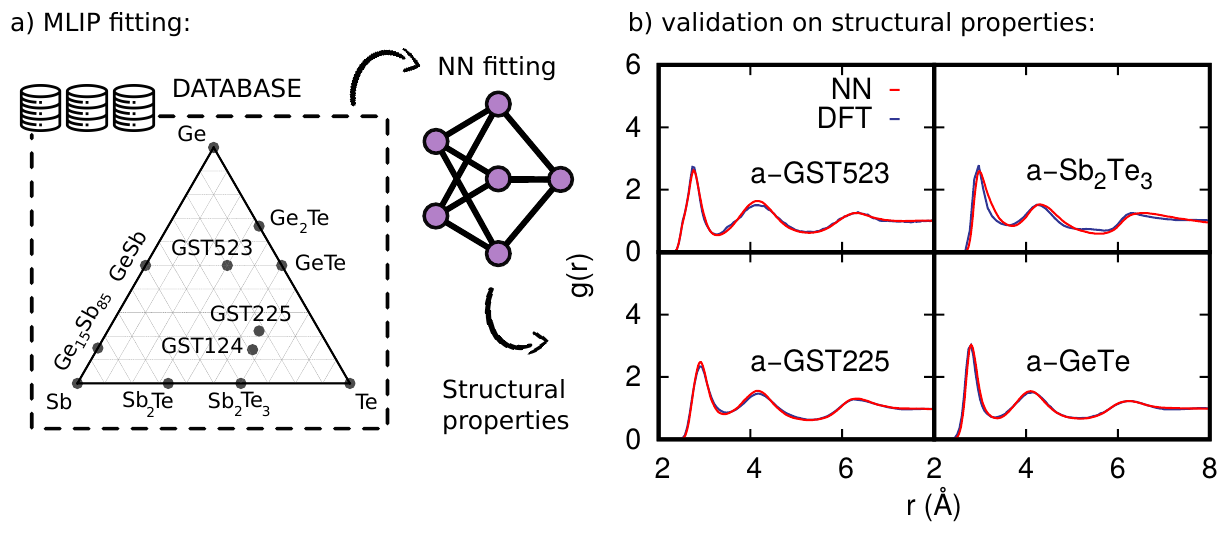}
     \caption{a) The ternary Ge-Sb-Te phase diagram showing the systems (dots) included in the training database of the MLIP. b) Total pair correlation  functions of amorphous GST523, GST225, Sb$_2$Te$_3$ and GeTe from  MLIP (red curves) and  DFT  (blue curves) simulations at 300 K. Details on these simulations are given in Sec. 2 of the SI.}
    \label{fig:validation}
\end{figure*}

\noindent
We generated the MLIP for  GGST alloys by fitting a DB of DFT energies, forces, and virial tensors within the neural network (NN) framework implemented in the DeePMD code \cite{{DeePMD4,DeePMD2,DeePMD3}}. To this end, we exploited the DB generated previously for GST225 compound and Ge-rich Ge$_x$Te alloys, and extended it to a wider range of compositions in the Ge-Sb-Te ternary phase diagram. The final DB consists of about 470,000  configurations of small cells in the amorphous and liquid phases and in the crystalline phases for the elemental and stoichiometric compounds that include
1) elemental Ge, Sb and Te,
2) the stoichiometric  GeTe, Sb$_2$Te and Sb$_2$Te$_3$ binary compounds,
3) the off-stoichiometric  Ge$_2$Te, Ge$_{15}$Sb$_{85}$ and GeSb binary alloys, 
4) the stoichiometric GST225 and GeSb$_2$Te$_4$ ternary compounds, 
5) the off-stoichiometric Ge-rich GST523 alloy,
6) crystalline superlattices GeTe-GST225, Sb$_2$Te$_3$-GST225, GST326-GST124.
The compositions included in the DB are highlighted in the ternary Ge-Sb-Te phase diagram in Fig. \ref{fig:validation}a.
The number of configurations and the cell sizes of each system are reported in Table S1 in the SI.\\

The atomic configurations were extracted from DFT molecular dynamics (MD) simulations by using the CP2k code \cite{quickstep2}.  The Perdew-Burke-Ernzherof (PBE) exchange and correlation functional \cite{PBE} and Goedecker-Teter-Hutter (GTH) norm conserving  pseudopotentials with $s$ and $p$ valence electrons were used \cite{GTH1,GTH2}.  Kohn-Sham orbitals were expanded in a Triple-Zeta-Valence-plus-Polarization (TZVP) basis set while the electronic density was expanded in plane waves up to a kinetic cutoff of 100 Ry. The Brillouin Zone (BZ) integration was restricted to the $\Gamma$-point in all MD simulations. The time step was set to 2 fs and configurations were extracted every 100-200 fs. We recalculated energies, forces, and stresses for each configuration added to the DB with a higher accuracy by increasing the kinetic energy cutoff for the plane-waves expansion of the electronic density to 400 Ry and by using a 3x3x3 or 4x4x4 k-point mesh for the BZ integration. A Fermi-Dirac smearing in the occupation of Kohn-Sham states was used with an electronic temperature of 300 K.\\

In NN schemes for the generation of interatomic potentials, the total energy of the system is written as the sum of individual atomic energies that depend on the local environment of each atom \cite{Behler}.
In the DeePMD scheme the local environment is encoded by local descriptors which are generated by an embedded neural network. A second neural network  (fitting network)  is built for the calculation of energies and forces with the local descriptors as input layer. We designed the embedded network with three hidden layers of 20, 40, and 80 nodes supplemented by other two layers with attention-based mechanism whose length is 128. The cutoff radius r$_c$ was set to 7 \r{A}, which is beyond the third coordination shell of our systems, while the smoothing radius r$_s$ was set to 0.5 \r{A} in the implementation of Ref. \cite{DeePMD2}. The maximum number of neighbors was set to 79. Finally, the network for the fitting of energy and forces consists of 4 hidden layers with 240 nodes each. In the embedding and fitting network, we have used the hyperbolic tangent as an activation function.

The MLIP was generated in an iterative manner. A first MLIP was generated by training over the DB of GST225 and Ge$_x$Te, supplemented with about 5000 configurations of GST523. Then, the MLIP is used to perform MD simulations that provide configurations from unexplored region of the configurational space to enrich the training DB for a second generation of the potential. In this step we performed metadynamics simulations \cite{laio,bonati2018silicon,niu2020ab} designed to accelerate the segregation of Ge in GGST alloys, and to collect configurations that include Ge-GeSbTe interfaces. We performed different simulations by using
as collective variable one of the partial coordination numbers Ge-Ge, Ge-Sb or Ge-Te to promote the formation of Ge-rich regions in contact with Ge-poor regions.
This step was crucial to accurately describe the interface energies and, hence, the barriers for the phase separation of GGST alloys.  At the same time we enriched the DB with additional atomic configurations extracted from DFT-MD simulations with different compositions, at different densities and exploring a wider temperature range. We iterated the procedure until the error on forces and energies did not improve any more nor the validation on the structural properties of the liquid, amorphous and crystalline phases improved further with respect to the very good results that we will illustrate in Section \ref{sec:validation}. Structural and dynamical properties obtained from MLIP simulations were  compared with results from DFT-MD simulations  performed with the CP2k code with the same parameters given above. MLIP-MD simulations  were performed by using the LAMMPS code  \cite{LAMMPS} as MD driver with the DeePMD plugin.\\

To assess whether an atom is crystalline in the simulations of the crystallization process,  we used the Steinhardt order parameters Q$^{dot}_{4}$ \cite{q4,q4bis} defined for each atom $i$  by
\begin{equation}
     Q^{dot}_{4}=\frac{1}{N_i}\sqrt{\sum_{j=1}^{N_i} \sum_{m=-4}^{4}q_{4m,i} q^*_{4m,j}}  \; \; ; \; \;  q_{4m,i}=\frac{1}{N_i} \sum_{j=1}^{N_i}Y_{4m}\left(\mathbf{\hat{r}}_{ij}\right),
\end{equation}

where $N_i$ in the number of neighbors of atom $i$ up to a given cutoff, $j$ is the neighbors index, $Y_{4m}$ is the $4^{th}$  spherical harmonic with degree $m$, and $ \mathbf{\hat{r}}_{ij}$ is the unit vector connecting the two atoms. A threshold  of Q$^{dot}_{4}$ = 0.80 was chosen for an atom to be crystalline.\\

To identify the products of the decomposition process of GGST we used the
Smooth Overlap of Atomic Position similarity kernel (SOAP). 
In the SOAP formalism, the local atomic density around each atom $j$  is expressed as a sum of Gaussian functions (here with broadening $\sigma_{at}$= 0.5 \r{A}) centered on the position of its  neighbors up to a given cutoff (here 9-12 \r{A}). Then, the density around atom $j$ is expanded in spherical harmonics and radial basis functions $g_{b}(|\mathbf{r}|)$ as $\rho_j (\mathbf{r}) = \sum_{blm} c_{blm} g_{b}(|\mathbf{r}|) Y_{lm}(\mathbf{\hat{r}})$. The coefficients of this expansion define  the so-called power spectrum matrix $p(j)_{b_1 b_2 l} = \pi \sqrt{8/(2l+1)}\sum_m (c_{b_1lm})^*  c_{b_2lm}$ whose elements are turned into a vector $\mathbf{p}_j$ which is calculated for each atom for several configurations along the MD trajectories.  
Then, we performed a principal component analysis (PCA) \cite{PCA} over the set of vectors $\mathbf{p}_j$, which were then clustered in the PCA space with the k-means algorithm \cite{kmeans}. Finally, we labeled each atom  according to the cluster index, which allowed us to define distinct chemical and structural environments and their evolution in time directly from the MD trajectories. The SOAP kernel was computed using the DScribe Python package \cite{dscribe2,dscribe} and the ASE Python library \cite{ase} while for the PCA and the clustering analysis with k-means algorithm, we used the scikit-learn python package\cite{scikit-learn}.

\section{Results and discussion}
\subsection{Validation of the potential}\label{sec:validation}

The accuracy of the MLIP is assessed by the Root Mean Square Error (RMSE) on energies and forces reported for each system in our DB (see table S1 in the Supplementary Information (SI)). The position in the ternary Ge-Sb-Te phase diagram of the systems   included in the training DB is shown in Fig. \ref{fig:validation}a.  The average RMSE of the MLIP over  the whole training data set is 7.62 meV/atom on energy and  148 meV/{\AA } on forces which is very close to the RMSE on the test data set of 7.7 meV/atom  and 148 meV/\AA. 
Typical average error obtained with DeePMD for highly disordered multicomponent systems are in the range 2-7 meV/atom and 90-145 meV/\AA  \cite{RMSE1,RMSE2,RMSE3,RMSE4}

\begin{figure*}[]
    \centering
    \includegraphics[width=0.8\linewidth,keepaspectratio]{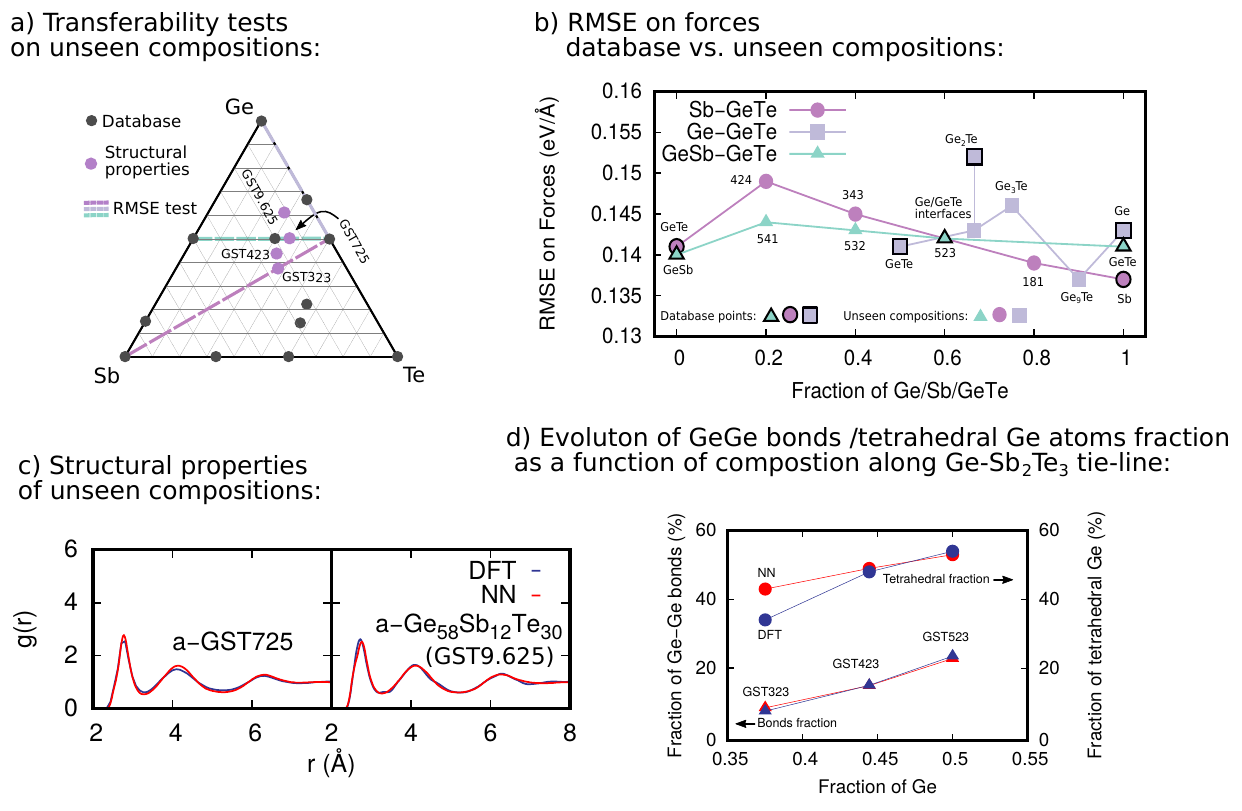}
     \caption{a) The Ge-Sb-Te ternary phase diagram with the systems included in the transferability test of the MLIP. The systems in the training DB are depicted with black dots. 
    A few compositions   not present in the training DB along the dashed lines were used as a test set of the MLIP by computing the RMSE on energies and forces. For this analysis, we considered alloys along the Sb-GeTe line (Sb-rich), Ge-GeTe (Ge-rich) and GeSb-GeTe line (both Sb-rich and Ge-rich). Finally,   the violet points indicate alloys not seen in the training on which the MLIP was validated by computing the structural properties of the amorphous phase  (Ge$_{9.{\bar 6}}$Sb$_2$Te$_5$, GST725, GST423, GST323). b) The RMSE on forces  for unseen compositions along the three lines in panel a), and for alloys in the DB (symbols with black border).  c) Total pair correlation function of amorphous GST723 and Ge$_{9.{\bar 6}}$Sb$_2$Te$_5$ from MLIP (red curves) and DFT (blue curves) simulations. Simulation details are given in Sec. 3 of the SI. d) Fraction (\%) of Ge atoms in tetrahedral geometry (dots) and of Ge-Ge homopolar bonds over the total number of bonds (triangles) as a function of Ge content (\%) in different alloys along the Ge-Sb$_2$Te$_3$ tie-line from MLIP (red) and DFT (blue) simulations. DFT data are taken from Ref. \cite{Omar2021}.  Simulation details are given in Sec. 3 of the SI.}
    \label{fig:transfer}
\end{figure*}

\begin{figure*}[]
    \centering
    \includegraphics[width=\linewidth,keepaspectratio]{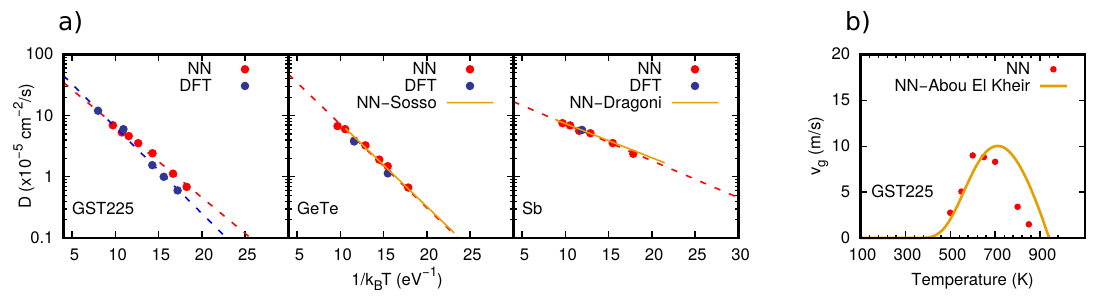}
     \caption{a) The Arrhenius plot (red  dashed lines and red dots) of the self-diffusion coefficient $D$ for GST225, GeTe and Sb. Data on GST225 are compared with previous DFT results from Ref. \cite{caravatiDIFF} (blue dots and dashed line). Data on GeTe are compared  with DFT results at two temperatures (blue dots) while the Arrhenius fit (red dashed line) is compared to previous results from NN simulations (orange continuous line) by Sosso {\sl et al.} \cite{sosso2012breakdown}. Data on Sb are compared at one temperatures with DFT results (blue dot) from Ref. \cite{bryk2023quasi} while the Arrhenius fit (red dashed line) is compared to previous results from NN simulations (orange continuous line) by Dragoni {\sl et al.} \cite{dragoni2021mechanism}. b) Crystal growth velocities ($v_g$) from present NN-MD simulations of GST225 compared with previous NN-MD simulations with a MLIP trained on GST225  from Abou El Kheir {\sl et al.} \cite{npjOmar}. The continuous orange lines refer to the fitting of NN-MD results on $v_g$ from previous works with the Wilson-Frenkel phenomenological formula (see Ref. \cite{npjOmar}).}
    \label{fig:msd_vg}
\end{figure*}

The accuracy of the MLIP is further assessed by comparing the structural properties of the liquid, amorphous and crystalline phases of the different systems included in the database  with reference DFT calculations. As an example, we show the total pair correlation function of 
amorphous GST225, GST523, Sb$_2$Te$_3$ and GeTe in Fig. \ref{fig:validation}b.
A  comprehensive analysis of the structural properties of all systems in the training DB is provided in Sec. 2 of the SI, which includes the partial pair correlation functions, the distribution of coordination numbers, the average partial coordination numbers, the angle distribution functions for each chemical species in amorphous and liquid phases. Moreover, it is well known that Ge in amorphous GST alloys can features different local environments: a tetrahedral geometry (4-fold coordinated), a pyramidal geometry (3-fold coordinated) and defective octahedral geometry (octahedral angles but coordination lower than six). The fraction of Ge atoms in tetrahedral configurations can be quantified by using the $q$-parameter for tetrahedricity introduced in Ref. \cite{qparam} and used for GST225 in Ref. \cite{caravati2007coexistence}. The $q$-parameter  is  reported for the all amorphous  alloys in the SI. We assessed the accuracy of the MLIP for the  crystalline phases by computing the  equation of state (energy as a function of volume) and its fit with the Birch–Murnaghan function which yields the equilibrium energy and density and the bulk modulus. 

Overall, the MLIP describes well all phases of all compositions. The largest error is on the equilibrium density of crystalline Te and Sb$_2$Te$_3$ which is overestimated by about 2 and 1.5 \% compared to reference data.

Since we are interested in reproducing the crystallization kinetics of GGST alloys which involves Ge segregation, the MLIP must be able to describe not only the starting and final compositions, but also all the intermediate compositions visited during the transformation process.  
Therefore, the MLIP must be transferable to a wide range of compositions around those present in the training DB.
The MLIP  was then cross-validated by evaluating the RMSE on compositions not seen during training. In this test set, we included compositions on the GeSb-GeTe line, Sb-GeTe line and configurations of the a-Ge/a-GeTe interface (see Fig. \ref{fig:transfer}a-b). The average RMSE on energies and forces, reported in Table S2 in the SI, is very similar to the average RMSE on the training database.

Moreover, we assessed the transferability of the MLIP by  evaluating the structural properties of the amorphous phase of four GGST alloys. Two alloys along the Ge-GST225 tie-line, namely Ge$_7$Sb$_2$Te$_5$ (GST725, or Ge$_{36}$-GST in the notation of Ref. \cite{Prazakova} which stands for 36 \% of Ge and 64 \% of GST225, which means a total of 36 + 64$\times$2/9 \% of  Ge) and Ge$_{9.{\bar 6}}$Sb$_2$Te$_5$ (Ge$_{58}$Sb$_{12}$Te$_{30}$, or Ge$_{46}$-GST in the notation of Ref. \cite{Prazakova}). The Ge$_{9.{\bar 6}}$Sb$_2$Te$_5$ alloy is close to the intersection between the Ge-GST225 tie-line and the GST523-Ge$_2$Te line. 
The other two alloys Ge$_4$Sb$_2$Te$_3$ (GST423) and Ge$_3$Sb$_2$Te$_3$ (GST323) lye along the Ge-Sb$_{2}$Te$_{3}$ tie-line. As an example here, in Fig. \ref{fig:transfer}c we compare the total pair correlation functions from NN and DFT simulations for GST725 and Ge$_{9.{\bar 6}}$Sb$_2$Te$_5$. 
Other  structural properties are discussed in Sec. 3 of the SI. Regarding the Ge-Sb$_{2}$Te$_{3}$ tie-line, we report the fraction of Ge-Ge homopolar bonds and the fraction of Ge in tetrahedral geometry in the amorphous phase as a function of Ge fraction in the alloy (see Fig. \ref{fig:transfer}d where NN data are compared with DFT data from Ref. \cite{Omar2021}). For the generation of these latter amorphous models we included the semiempirical van der Waals (vdW) corrections D3 \cite{D3} for consistency with the DFT calculations in Ref. \cite{Omar2021} we compare with. Other structural properties of these amorphous alloys are discussed Sec. 3 of the SI. \\

Overall, the MLIP reproduces well the reference data of energies and forces in  both the training and  transferability set. The MLIP describes very well the structural properties of all systems in the DB and of several alloys outside the training DB.
In particular, the MLIP is able to reproduce the structural properties of alloys along the three lines GST523-Sb$_2$Te$_3$, GST523-GeTe and GST523-Ge$_2$Te whose end points are all included in the training DB.
\\

Finally, we assessed the ability of the MLIP in reproducing the dynamical properties in the liquid and supercooled liquid phases. 
The self-diffusion coefficient $D$ as a function of temperature is shown in Fig. 
\ref{fig:msd_vg}a for GST225, GeTe and Sb as obtained from the present NN-MD simulations and from previous DFT-MD or NN-MD simulations.  $D$  is computed  from the mean square displacement (MSD) and  the Einstein relation MSD = $6Dt$ at long time $t$.  In the temperature range of Fig. \ref{fig:msd_vg}a, $D$  can be described by an Arrhenius function $D$=$D_0 exp(-E_a/k_BT)$. The activation energy $E_a$ for GST225, GeTe and Sb from NN-MD  (reference data) simulations is 0.270 (0.333)\cite{caravatiDIFF}, 0.310 (0.296)\cite{SossoNN} and 0.138 (0.130)\cite{dragoni2021mechanism} eV. 
In supercooled liquid GST523 at 835 K the NN-MD self-diffusion coefficient of 4 $\times$ 10$^{-5}$ cm$^2$/s is very close to the  result of 
3.8 $\times$ 10$^{-5}$ cm$^2$/s from previous DFT-MD simulations \cite{Omar2021}. In these last simulations on GST523, we used the D3  vdW corrections \cite{D3} for consistency with the previous DFT work we compare with \cite{Omar2021}.
Therefore, the MLIP reproduces well also previous  results in the literature on the self-diffusion coefficient from DFT-PBE or NN-MD simulations fitted on DFT-PBE data.

 We took a step further by simulating the heterogeneous   crystallization process in GST225.
 The crystal growth velocities from the present NN-MD simulations are compared with NN-MD results from previous works in Fig. \ref{fig:msd_vg}b.
 We remark that with our MLIP the melting temperature T$_{\rm m}$  for GST225 is 872 K, which is slightly closer to the experimental value of 900 K
 \cite{mazzaGeTeliquid} than the value of 940 K obtained with the previous MLIP for GST225 of Ref. \cite{npjOmar}.
 To estimate T$_{\rm m}$ we used the same phase coexistence protocol described in our previous work \cite{npjOmar}.

We remark that so far we have validated the MLIP  by comparison to DFT-PBE data and to results from previous NN-MD simulations with potentials fitted on DFT-PBE data. The reliability of the DFT-PBE framework in describing the  experimental structural and dynamical properties of phase change alloys  and  in particular GST225 has been discussed  in several previous works \cite{caravati2007coexistence,akola2007structural,elliot225,zhang2019designing,acharya2025simulation} to which we refer to for a comprehensive analysis. We just mention that by adding vdW D2 corrections to our previous MLIP potential for GST225 \cite{npjOmar} a very good agreement with experiments is achieved on the crystal growth velocities  
\cite{acharya2025simulation} and on the viscosity of the liquid \cite{marcorini}.

In conclusions, we judge that our  MLIP is sufficiently accurate to investigate the kinetics of crystallization and phase separation in GGST that is discussed in the next section.

\section{Simulation of the decomposition and crystallization process of Ge-rich GST alloys} 
The MLIP was exploited to simulate the crystallization and phase separation process of GGST alloys. We considered  the three compositions already discussed in Sec. \ref{sec:validation}, namely GST523  on the Ge-Sb$_2$Te$_3$ tie-line  and GST725 on the Ge-GST225 tie-line which have the same Ge fraction (50 \%). The third system is the  Ge$_{9.{\bar 6}}$Sb$_2$Te$_5$ alloy (Ge$_{58}$Sb$_{12}$Te$_{30}$, Ge$_{46}$-GST in the notation of Ref. \cite{Prazakova}) that is richer in Ge (58 \%) and lies 
on the Ge-GST225 tie-line.  All simulations were performed at 600 K and at constant volume, consistent with the actual operation conditions of the memory.

As we mentioned in section \ref{sec:intro}, from a thermodynamical point of view, GGST alloys are supposed to decompose into Ge and stoichiometric GST compounds on the Sb$_2$Te$_3$-GeTe tie-line. However, only the first stages of the transformation might be completed on the short time span of the programming operations of the memory, and kinetic effects could give rise to metastable intermediate phases as discussed in the introduction. 

On these premises, we investigated the crystallization with phase separation  aiming at identifying the intermediate products that form on the ns time scale at 600 K.

In the case of GST523, we used a 14080-atom supercell at the theoretical equilibrium density of amorphous GST523 (0.0338 atom/\AA$^3$)\cite{ghezzi} obtained in our previous work \cite{Omar2021}. Assuming that at the very end, GST523 decomposes into the thermodynamically favored products (Ge and Sb$_2$Te$_3$), and that Ge segregates into the amorphous phase at its equilibrium density (0.0438 atom/\AA$^3$)\cite{baribeau}, the  density of the remaining Sb$_2$Te$_3$-like region would be 0.0275 atom/\AA$^3$, which is very close to the experimental density of the liquid at 1003 K (0.0281 atom/\AA$^3$) \cite{singh2003structural}. We started from a model equilibrated at 1400 K for 20 ps, followed by a second equilibration at 1000 K for 20 ps, and then by a rapid quenching to 600 K in 40 ps. Once we reached the target temperature,  we simulated the crystallization with phase separation for about 6 ns at NVT conditions. The SOAP analysis described in Sec. \ref{sec:methods} was then performed over the whole trajectory.
The cluster analysis (see Sec. \ref{sec:methods}) yielded ten different environments that have been identified by cross-checking with the distribution of the different chemical species. The ten environments are described in Table \ref{environments}.

\begin{table}
\begin{center}
\begin{tabular}{|c| c | }
\hline
Environment  & description \\ 
number &  \\
\hline

1 & Ge atoms at interfaces \\ 
&\\
2 &Te atoms in GeTe-like regions\\ 
&\\
3 &Ge atoms in GeTe-like regions\\ 
&\\
4 &Sb in GeSb-like regions in GST523,\\
& both Ge and Sb in GeSb-like regions for other alloys\\ 
&\\
5 & Sb in homogeneous GST\\ 
&\\
6 & Ge in homogeneous GST\\
&\\
7 & Sb atoms at interfaces \\
&\\
8 & Te in homogeneous GST  \\
&\\
9 & Ge in GeSb-like region in  GST523 and GST725,\\
& pure Ge atoms in  Ge$_{9.{\bar 6}}$Sb$_2$Te$_5$ \\
&\\
10 &Te atoms at interfaces \\
&(present only in  GST523) \\
&\\
\hline
\end{tabular}
\end{center}
\caption{Description of the ten different environments identified by the SOAP and cluster analysis (see Sec. \protect\ref{sec:methods}) during the transformation of the three alloys GST523, GST725 and Ge$_{9.{\bar 6}}$Sb$_2$Te$_5$ at 600 K.}
\label{environments}
\end{table}

\begin{figure}[h!]
    \centering
    \includegraphics[width=1.0\linewidth,keepaspectratio]{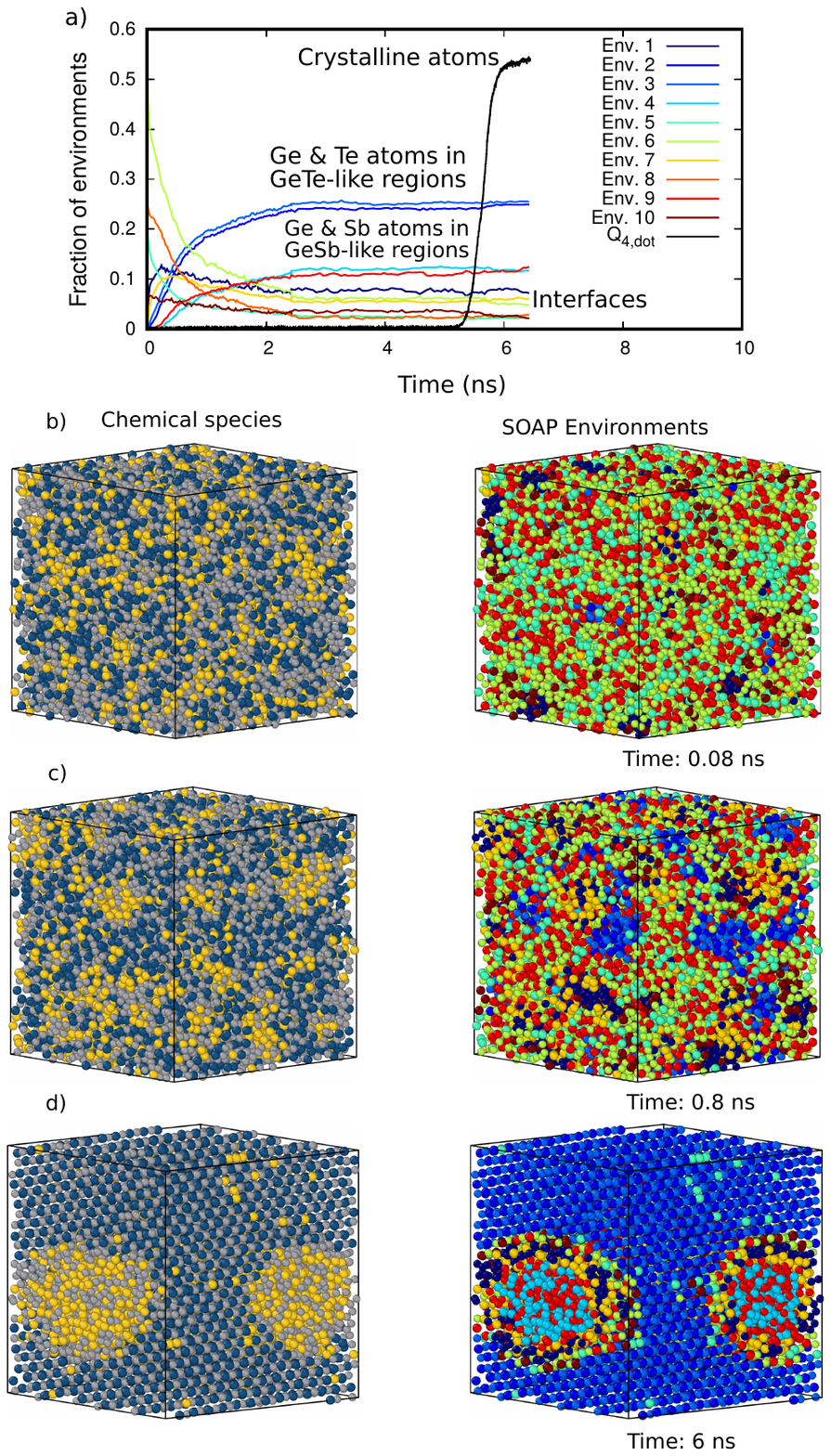}
     \caption{a) Evolution in time of the fraction of atoms in the ten different SOAP environments and of the  Q$^{dot}_{4}$ order parameter for crystallinity, during the crystallization of GST523. The description of the different environments is given in Table \ref{environments}. Snapshot of the transformation process  b) at the beginning, c) at an intermediate time and d) at the end of the simulation. In the panels on the left atoms are colored according to the chemical species (Ge, Sb, Te in gray, yellow and blue). In the panels on the right, atoms are colored according to the different SOAP environments whose color code is given in panel a). The Ovito \cite{ovito} tool was used for the visualization and the generation of all
atomic snapshots of this article.}
    \label{fig:crys523}
\end{figure}

The time evolution of the fraction of atoms belonging to the different ten SOAP environments is shown in Fig. \ref{fig:crys523}a together with the fraction of crystalline atoms identified by the Q$^{dot}_{4}$ order parameter for crystallinity (see Sec. \ref{sec:methods}). Snapshots at an early stage (0.08 ns), an intermediate stage (0.8 ns) and at the end of the simulation are shown in Fig. \ref{fig:crys523}b-c which reports  the atoms colored  according to the chemical species (on the left) and according to the SOAP environment (on the right).
At the beginning of the simulation (Fig. \ref{fig:crys523}b), the system is almost homogeneous and consists of Sb, Ge and Te in  homogeneous GST523 (SOAP environments 5,6,8, see Table \ref{environments}). In the intermediate and final snapshots (Fig. \ref{fig:crys523}c-d),  Te and Ge atoms in GeTe-like region are found (SOAP environments 2 and 3).  
The decomposition also leads to the appearance of  Sb and Ge atoms in GeSb-like regions (SOAP environments 4 and 9) and  atoms at the interface between GeTe-like and GeSb-like regions (SOAP environments 1, 7  and 10).\\

In the simulation,the phase separation starts almost immediately, as was the case for Ge$_2$Te \cite{baratella2025}. In the time evolution  in Fig. \ref{fig:crys523}, we first observed an increase in the fraction of interfacial Ge and Sb atoms (SOAP environments 1 and 7), followed by the formation of a GeTe-like region (SOAP environments 2 and 3). Subsequently,  GeSb-like regions appear (SOAP environments 4 and 9). After about 2 ns the transformation products reach a plateau. Once phase separation was completed, we observed a coarsening of the different GeSb regions formed during the previous steps, as shown  in Fig. \ref{fig:crys523}c-d (SOAP environments 4 and 9). Finally, a GeTe-like crystalline nucleus forms and grows after about 5 ns (black curve in Fig.\ref{fig:crys523}a) until it reaches a saturation corresponding to complete crystallization of the GeTe-like regions that percolate through the cell along the three cartesian directions (see Fig. \ref{fig:crys523}). The products of the crystallization and  phase separation of GST523 on the ns time scale  are thus crystalline Ge$_{12}$SbTe$_{12}$ (slightly Sb-doped GeTe) and amorphous Ge$_{52}$Sb$_{48}$, instead of the thermodynamically favored Ge and Sb$_2$Te$_3$ crystals. 
We remark that  at the end of the simulation  the cluster analysis assigns to environments 5,6, 8
(homogeneous untransformed GST) atoms that are actually at the interfaces between crystalline GeTe and amorphous GeSb/Ge but that clearly belong to one of these two regions. Therefore, the fraction of  residual  untransformed GST is smaller than the value given in Fig. \ref{fig:crys523}.

Very similar results were observed for the crystallization of GST725 on the Ge-GST225 tie-line (see Sec. 4 in SI for more details). In this case, the products of crystallization and phase separation is Ge$_{12}$SbTe$_{12}$ and amorphous Ge$_{54}$Sb$_{46}$ instead of Ge and GST225 as expected from thermodynamics. These simulations suggest that the transformation of GST523 and GST725 on the ns time scale is controlled by kinetics.

Switching to crystallization with phase separation of Ge$_{9.{\bar 6}}$Sb$_2$Te$_5$, we used an 8000-atom simulation box at the theoretical density of the liquid phase at 1000 K estimated form NPT simulation (0.03359 atom/\AA $^3$, See section 3 of SI). Similarly to GST523, by assuming that Ge$_{9.{\bar 6}}$Sb$_2$Te$_5$  will decompose into the thermodynamically favored products (Ge and GST225), and that Ge will segregate into the amorphous phase at its equilibrium density, the density of the remaining GST225 is 0.03111 atom/\AA$^3$ which is very close to the experimental density of the amorphous phase (0.0309 atom/\AA$^3$). We followed the same protocol used for GST523, by quenching from 1400 K to 600 K, followed by about 3.2 ns simulation of the crystallization with phase separation. Then, we performed the SOAP analysis as described above for GST523. The cluster analysis (see Sec. \ref{sec:methods}) yielded the same 
environments identified in GST523 (see Table \ref{environments}), but for interfacial Te atoms that were not found in this latter alloy richer in Ge.

\begin{figure}[h!]
    \centering
    \includegraphics[width=0.8\linewidth,keepaspectratio]{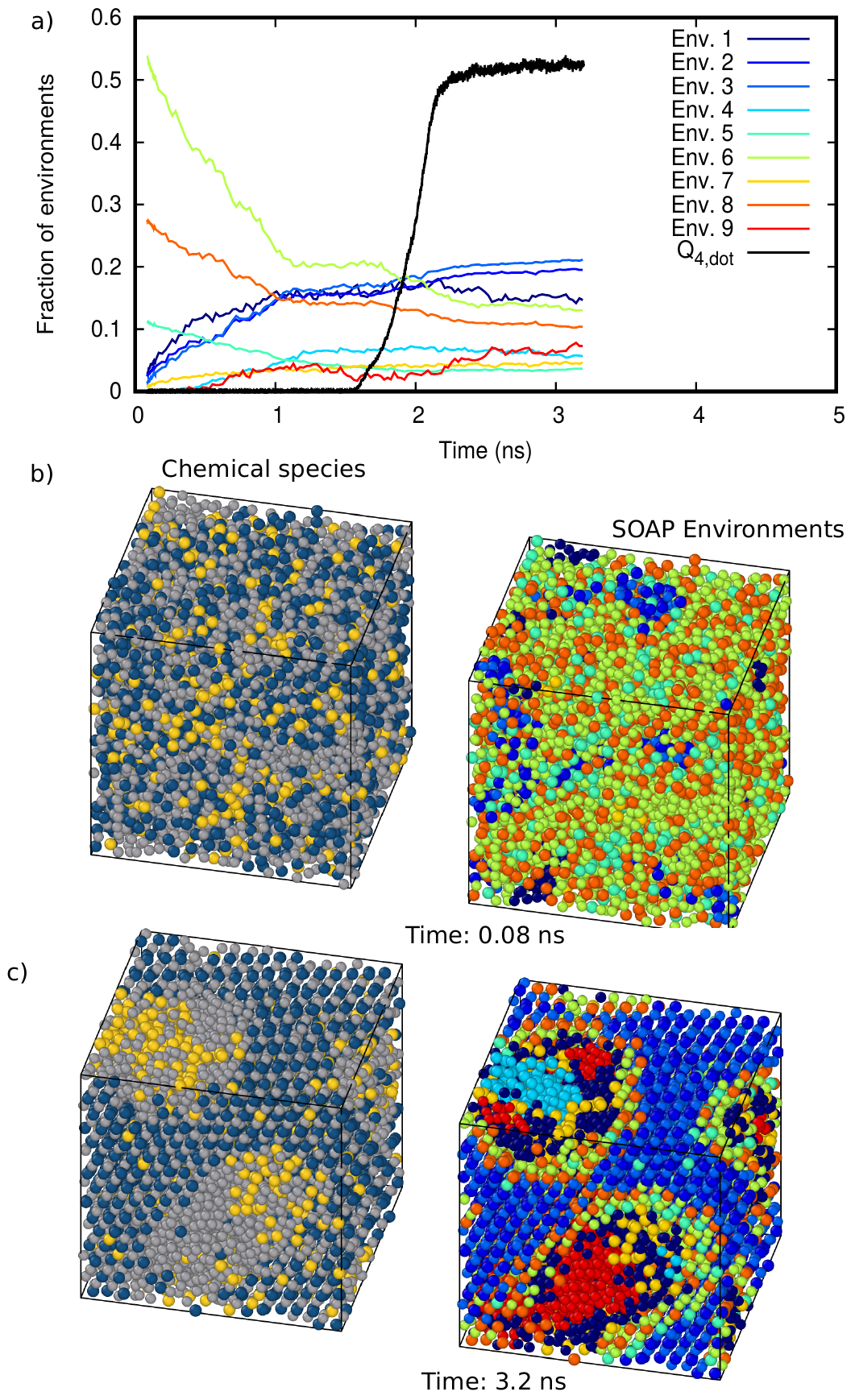}
     \caption{a) Evolution in time of the fraction of atoms in the nine different SOAP environments and of the Q$^{dot}_{4}$ order parameter for crystallinity,  during the crystallization of Ge$_{9.{\bar 6}}$Sb$_2$Te$_5$. The description of the different environments is given in Table \ref{environments}. Snapshot of the transformation process  b) at the beginning and c) at the end of the simulation. In the panels on the left atoms are colored according to the chemical species (Ge, Sb, Te in gray, yellow and blue). In the panels on the right, atoms are colored according to the different SOAP environments whose color code is given in panel a).}
    \label{fig:crys581230}
\end{figure}

The time evolution of the fraction of atoms belonging to the different nine SOAP environments is shown in Fig. \ref{fig:crys581230}a together with the fraction of crystalline atoms identified by the Q$^{dot}_{4}$ order parameter for crystallinity (see Sec. \ref{sec:methods}). Snapshots at an early stage (0.08 ns) and at the end of the simulation are shown in Fig. \ref{fig:crys581230}b-c which reports atoms colored according to the chemical species (on the left) and according to the SOAP environment (on the right).
The results are very similar to those of GST523 and GST725.
First, a separation into amorphous (supercooled liquid)  GeTe-like and Ge-Sb-like regions is observed and then, after the phase separation was completed, crystal nucleation and growth was observed within the GeTe-like regions. A slightly Sb-doped cubic GeTe crystal percolates through the simulation cell along the three directions (see Fig. \ref{fig:crys581230}b-c). As opposed to GST523 and GST725, for the more Ge-rich Ge$_{9.{\bar 6}}$Sb$_2$Te$_5$ alloy, we also observed regions of pure amorphous Ge.

The products of the crystallization and phase separation of Ge$_{9.{\bar 6}}$Sb$_2$Te$_5$ on the ns time scale  are finally crystalline Ge$_{48}$Sb$_5$Te$_{47}$, two pure Ge clusters  and two amorphous Ge$_{46}$Sb$_{54}$ clusters, instead of the thermodynamically favored Ge and GST225 crystals.  

The simulations of all three compositions show that at 600 K and on the time scale of a few ns, which are comparable to the operation conditions of the set process in the memory, the GGST alloy decomposes into amorphous GeSb and slightly Sb-doped GeTe regions. For the more rich Ge$_{9.{\bar 6}}$Sb$_2$Te$_5$ alloys, also regions of pure amorphous Ge are observed.  After completion of the phase separation, crystal nucleation takes place in the GeTe-like regions and the crystal growth of cubic GeTe proceeds percolating through the whole cell, while the GeSb regions remain amorphous. These phases are  intermediate products that differ from the thermodynamically stable phases that should form according to the convex hull construction, namely pure Ge and GST225 or Sb$_2$Te$_3$. Only on a longer time scale, not accessible by the NN simulations, we expected to see a Ge/Sb exchange between the GeSb and GeTe   regions to finally yield the thermodynamically stable products. 
The intermediate phases we see should correspond to the product of the set process in the memory. Indeed GeTe-like crystallites, possibly Sb-doped,  were identified  in TEM and EELS measurements in GGST memory cells in Ref. \cite{petroni2025}. We remark that diffraction data cannot easily  discriminate between Sb-doped GeTe and the stoichiometric  GST alloys on the GeTe-Sb$_2$Te$_3$ tie-line due to the similarity of their lattice constants. On the other hand,  EELS and dispersive X-ray measurements can 
provide only a composition averaged over different grains along the direction of the beam which introduces uncertainties in the 
measured composition of the crystallites. The present simulations thus provide an important support to the interpretation of the experimental characterization of the operation of the memory cell.  We also remark that the formation of GeTe-like crystal in the decomposition of GGST was also predicted by thermochemical data based on DFT calculations in Ref. \cite{OmarHigh}. It was shown, that besides the thermodynamically stable compounds on the GeTe-Sb$_2$Te$_3$ tie-line, Sb-doped GeTe crystals in a large region of the phase diagram close to pure GeTe are highly favored in the decomposition of GST523 \cite{Yimam2022} and GST725 \cite{cecchi2022crystallization} because they lie close to the convex hull \cite{OmarHigh}. The formation of cubic GeTe-like crystals was also observed in the crystallization of thin films with composition along the Ge-GST225 tie-line \cite{privitera2020crystallization, rahier}, and 
of two alloys with composition close to GST523 (at 300  $^o$C) and GST725 (at 330  $^o$C) \cite{cecchi2022crystallization}. 
In another experimental work, the decomposition of GST523 at 450 $^o$C was shown to yield GST213 and GST324 alloys \cite{Yimam2022}, which again can be seen an intermediate metastable phases toward the formation of the thermodynamically stable  Ge and Sb$_2$Te$_3$. 
On the contrary, we do not see the $Pnma$ phase of GeTe found in Ref. \cite{rahier} 
in the crystallization of Ge$_{0.5}$-GST, but at a lower temperature and on the time scale of hours which are far from the operation conditions of the memory that we addressed in our simulations. In this respect, we mention that the nature of the initial stages of phase separation and crystallization can depend on temperature. In our previous work on Ge$_2$Te we showed that at 600 K, segregation of most of Ge in excess was observed to precede the  crystallization of nearly stoichiometric GeTe regions, while at 500 K nucleation
of crystalline GeTe was observed to occur before phase separation, followed by a slow crystal growth due to
the concurrent expulsion of Ge in excess. We found a similar behavior in GST725 at 500 K, in which  the nucleation of GeTe-like crystals  preceded phase separation, and then a much slower crystal growth was observed with the expulsion of Ge in excess. It is therefore conceivable that on a much longer time scale and at other temperatures, different intermediate products could form for some compositions, as observed experimentally. We remark once more that here we are interested in uncovering the kinetics of crystallization and phase separation on ns time scale of interest for the memory and not on the macroscopic time scale addressed in experiments on thin films. 

\section{Conclusions}
In summary, we developed a MLIP for GGST alloys of interest for applications in  phase change memories embedded in microcontrollers. The MLIP was generated with the DeePMD code by fitting a large database of DFT energies and forces of elemental, binary, stoichiometric ternary compounds in the Ge-Sb-Te ternary phase diagram and the GST523 non-stoichiometric Ge-rich alloy. The MLIP is highly transferable in large regions of the phase diagram around the compositions included in the dataset. In particular, we demonstrated the transferability to  compositions on the Ge-GST225 and Ge-Sb$_2$Te$_3$ tie-lines. 
The MLIP was then exploited to simulate the crystallization with phase separation of the three compositions GST523, GST725 and Ge$_{9.{\bar 6}}$Sb$_2$Te$_5$, at 600 K and on the ns time scale which are comparable to the operation conditions of the memories.
For all compositions, phase separation was completed in 1-3 ns yielding slightly Sb-doped GeTe  and GeSb amorphous regions. In the more Ge-rich Ge$_{9.{\bar 6}}$Sb$_2$Te$_5$ alloys also pure amorphous Ge regions were observed. Then, crystal nucleation and growth take place in the GeTe-like region giving rise to a crystal percolating through the cell. The final products consisting of cubic Sb-doped GeTe and amorphous GeSb/Ge are intermediate metastable phases that form due to kinetics effects, the thermodynamically stable phases being pure Ge and GST225 or Sb$_2$Te$_3$ that could form only in a much longer time scale. The cubic GeTe-like crystal is 
expected to form on the short time span and at operation temperature of the set process of the memory, as confirmed by recent TEM measurements \cite{petroni2025}. 
Other intermediate products such as Sb-rich regions and Sb crystals that have been identified in the memories by TEM measurements in Ref. \cite{petroni2025}, could result from the presence of a strong  electric field and an inhomogeneous temperature field that  might affect the decomposition process.
The MLIP  can be exploited also to simulate the set process under these more realistic conditions of the operation of the memory in very large models, as we recently did for stoichiometric GST225 \cite{npjOmar}.

\bibliography{biblio}

\vskip 1 truecm

\section*{Code availability}
LAMMPS, and DeePMD are free and open source codes available at https://lammps.sandia.gov and http://www.deepmd.org, respectively.

\section*{Data Availability Statement}
The NN potential,  the training DFT database, atomic trajectories of the phase separation and crystallization process, and a video of the relevant processes will be available in the Materials Cloud repository after acceptance of the article.

\section*{Author Contributions}
O. Abou El Kheir and M. Bernasconi contributed equally to the work. D. Baratella contributed to the development of the database and of the machine learning potential.

\section*{Conflicts of interest}
There are no conflicts of interest to declare.

\section*{Acknowledgments}
The project has received funding  from European Union Next-Generation-EU  through the Italian Ministry of University and Research under PNRR M4C2I1.4 ICSC  Centro Nazionale di Ricerca in High Performance Computing, Big Data and Quantum Computing (Grant No. CN00000013). 
We acknowledge the CINECA award under the ISCRA initiative, for the availability of high-performance computing resources and support.

\end{document}


\date{}
\maketitle
\tableofcontents

\newpage
\section{Database}
\begin{figure}[H]
    \centering
    \includegraphics[width=1\linewidth,keepaspectratio]{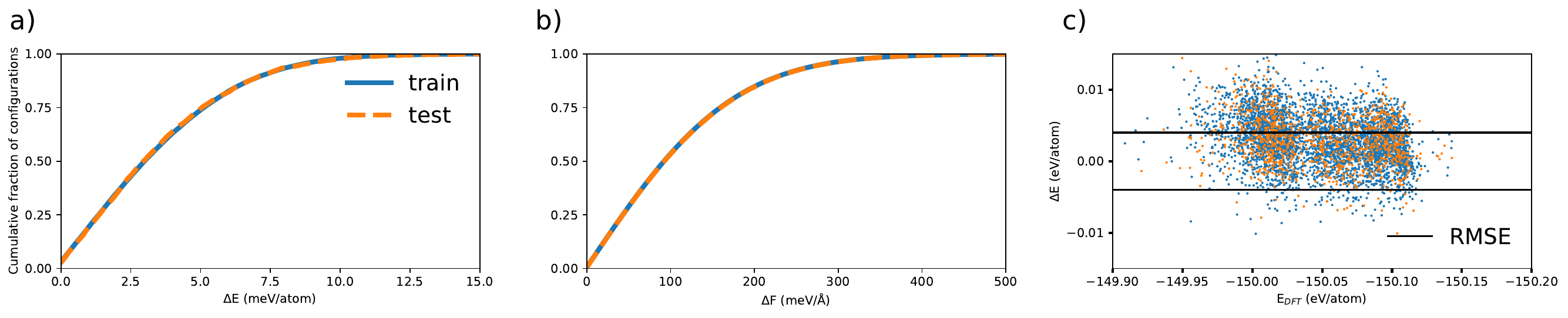}
    \caption{Cumulative fraction of the absolute errors of the NN potential in training and test data sets of GST523 (a) the energies per atom ($\Delta$E=|E$_{DFT}$-E$_{NN}$|) and (b) forces ($\Delta$F=|F$_{DFT}$-F$_{NN}$|). c) The distribution of the absolute error as a function of DFT energy.}
    \label{fig:distRMSE}
\end{figure}

In Table \ref{database} we report on the number of configurations for each compound/alloy in the database, the cell size, the Root Mean Square Error (RMSE) on energy and forces for  both  the training and test datasets.  Similarly, in Table \ref{database2} we report on the same information for the transferability test set on compositions not included in the training database. Moreover, in Fig. \ref{fig:distRMSE}a-b we report on the cumulative fraction of the absolute errors of the NN potential in training and test data sets of GST523 on the energies per atom and forces,  while in Fig. \ref{fig:distRMSE}c we report on the distribution of the absolute error as a function of DFT energy.
The training database conists of about 470,000  configurations of small cells in the amorphous and liquid phases and in the crystalline phases only for the elemental and stoichiometric compounds. The database for
GST523 also include configurations of Ge-GeSbTe interfaces generated by metadynamics simulations  designed to accelerate the segregation of Ge in GGST alloys (see article).

\begin{table} [H]
\centering
  \begin{adjustbox}{width=\textwidth}
\begin{tabular}{lccccc}
\hline\hline
Composition & Number of configurations & cell size (atoms)& RMSE on energy Training/Test (meV/atom) & RMSE on Forces Training/Test (eV/\AA) \\
\hline\hline
Ge                  &  39132&64-144 &  6.6 (6.6)       & 0.143 (0.141)\\
Sb                  &  19011&54-72  &  6.0 (5.9)       & 0.124 (0.123) \\
Te                  &  17354&64-162 &  13.3 (12.2)     & 0.200 (0.189)\\
GeSb                &  21005&64-72  &  5.6 (5.5)       & 0.132 (0.131) \\ 
Ge$_{15}$Sb$_{85}$  &  10042&64     &  6.1 (5.9)       & 0.128 (0.128)\\
GeTe                &  42900&96-108 &  7.8 (8.0)       & 0.141 (0.142) \\
Ge$_2$Te            &  51300&100    &  8.2 (8.2)       & 0.152 (0.152)\\
Sb$_2$Te            &   8819&72-81  &  5.9 (5.9)       & 0.161 (0.159)\\
Sb$_2$Te$_3$        &   9649&60-135 &  11.4 (11.5)     & 0.186 (0.188)\\
GeSb$_2$Te$_4$      &   7048&56     &  5.7 (5.9)       & 0.141 (0.141)\\
Ge$_2$Sb$_2$Te$_5$  & 183654&98-216 &  8.8 (9.2)       & 0.155 (0.156)\\
Ge$_5$Sb$_2$Te$_3$  &  47315&150    &  4.0 (4.0)       & 0.142 (0.142)\\
SL                  &  11700&68-76  &  2.2 (2.2)       & 0.084 (0.084)\\
\hline\hline

\end{tabular}
  \end{adjustbox}
\caption{Number of configurations included in the database for each composition, cell size and the average RMSE on energy and forces for training (test) subset. SL refer to superlattices of crystalline: GeTe-GST225, Sb$_2$Te$_3$-GST225 and GST326-GST124.   }
\label{database}
\end{table}

\begin{table} [H]
\centering
  \begin{adjustbox}{width=\textwidth}
\begin{tabular}{lccccc}
\hline\hline
Composition & Number of configurations & cell size (atoms)& RMSE on energy Test (meV/atom) & RMSE on Forces Test (eV/\AA) \\
\hline\hline
GeSb$_8$Te         &  5010 & 100 & 7.9  & 0.139\\
Ge$_3$Sb$_4$Te$_3$ &  5010 & 100 & 6.6  & 0.145\\
Ge$_4$Sb$_2$Te$_4$ &  5010 & 100 & 11.0 & 0.149\\
Ge$_5$Sb$_3$Te$_2$ &  5010 & 100 & 6.0  & 0.143\\
Ge$_5$Sb$_4$Te$_1$ &  5010 & 100 & 12.0 & 0.144\\
Ge$_{7.4}$Sb$_{1.3}$Te$_{1.3}$ &  5010  & 100 &  8.2 & 0.141\\
Ge$_2$Te interfaces & 800 & 300 & 8.83  & 0.143 \\
Ge$_3$Te            & 2000 & 300 & 13.5 & 0.146 \\
Ge$_9$Te            & 2000 & 300 & 10   & 0.137 \\
\hline\hline
\end{tabular}
  \end{adjustbox}
\caption{Number of configurations in the transferability data set for each composition, cell size and the average RMSE on energy and forces for test datasets. The configurations refer to the liquid and amorphous phases for all compositions. Ge$_2$Te interfaces refer to configurations of the interface between Ge$_2$Te and Ge taken from Ref. \cite{baratella2025}.}
\label{database2}
\end{table}

\section{Validation of the potential }
Hereafter, we report the validation of the NN potential on structural, dynamical and thermodynamical properties of GeSbTe alloys. To study the liquid phase,  we first prepared a model equilibrated at 2000 K for 20 ps for each alloy. Then, the models were quenched in 20 ps down to a target temperature  (T$_{target}$) in 30 ps and finally the models were equilibrated at the T$_{target}$ for 50 ps before evaluating  the average properties. T$_{target}$, in the range 1200-990 K, was chosen for each alloy to be higher than the melting temperature.

\subsection{$Ge_2Sb_2Te_5$}
\begin{figure}[H]
    \centering
    \includegraphics[width=0.8\linewidth,keepaspectratio]{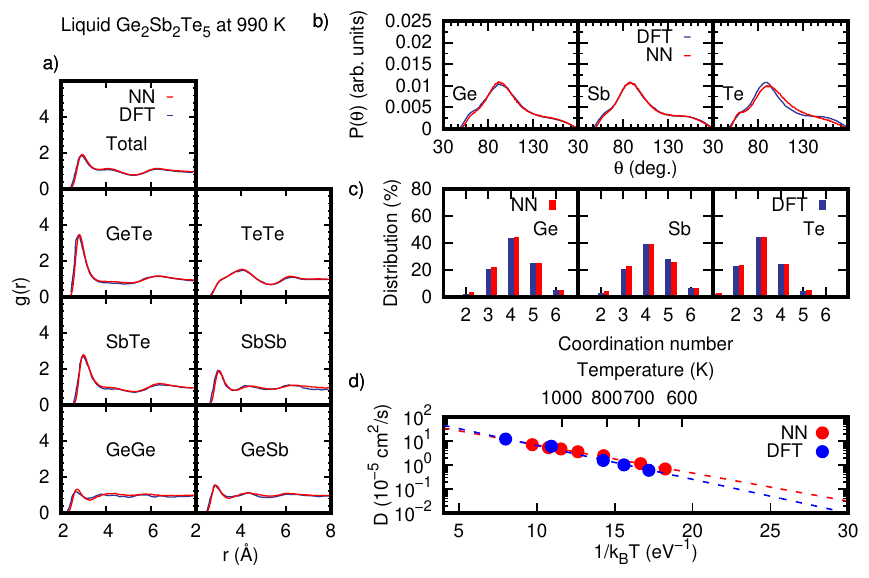}
    \caption{Structural and dynamical properties of liquid GST225 from DFT (blue curves) and NN (red curves) simulations with a 216-atom or 999-atom models, respectively. a) Total and partial radial distribution functions at 990 K. The position of the first maximum and minimum of each partial correlation function are given in Table \ref{tab:gdrmax225liquid}. b) Angle distribution function resolved per central atomic species at 990 K. The data were normalized to the number of triplets in each model. c) Distribution of coordination numbers resolved per chemical species at 990 K computed by integrating the partial pair correlation functions up to a bonding cutoff which corresponds to 3.2 \AA\  for all pairs except for Sb–Te for which we use 3.4 \AA. We use the same bonding cutoff of the amorphous phase. d) The Arrhenius plot of the diffusion coefficient D as a function of temperature compared with previous DFT results with the PBE functionals from Ref. \cite{npjOmar}}
    \label{fig:225liquid}
\end{figure}

\begin{figure}[H]
    \centering
    \includegraphics[width=0.8\linewidth,keepaspectratio]{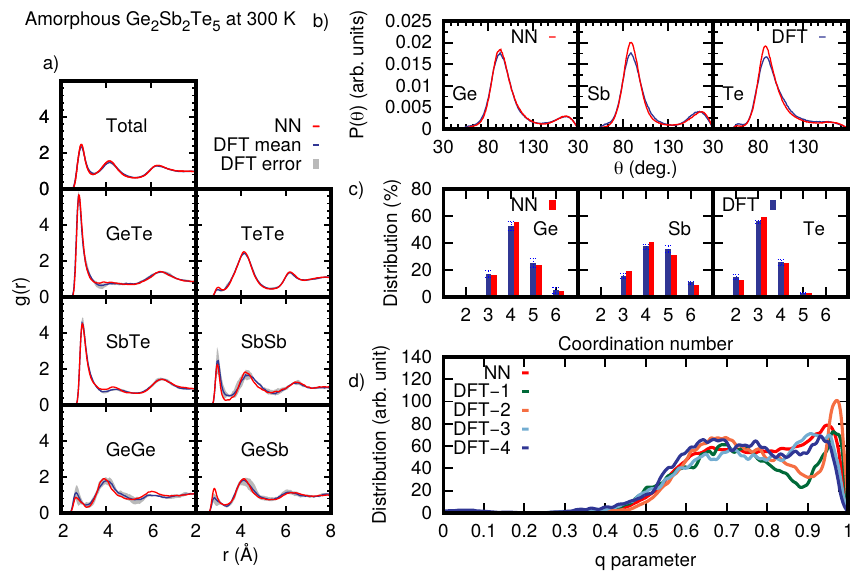}
    \caption{Structural properties of amorphous GST225 at 300 K from NN (red curves) and DFT (blue curves) simulations. DFT data are averaged over four independent 216-atom models while NN data are obtained from a 999-atom model. a) Total and partial radial distribution functions. The position of the first maximum and minimum of the different functions is reported in Supplementary Table \ref{tab:agdrmax}. b) Angular distribution function resolved per central atomic species. The data are normalized to the number of triplets in each model. c) Distribution of coordination numbers resolved per chemical species, obtained by integrating the pair correlation functions up to a bonding cutoff corresponding to 3.2 \AA\ for all pairs except for Sb-Te for which a longer cutoff value of 3.4 \AA\ was used. The spread over the four DFT models are indicated by error bars. d) Distribution of the local order parameter $q$ for tetrahedricity for four-coordinated Ge atoms for the NN model and for each of our four DFT models.}
    \label{fig:225amorphous}
\end{figure}

\begin{figure}[H]
 \centering
 \includegraphics[width=0.6\textwidth, keepaspectratio]{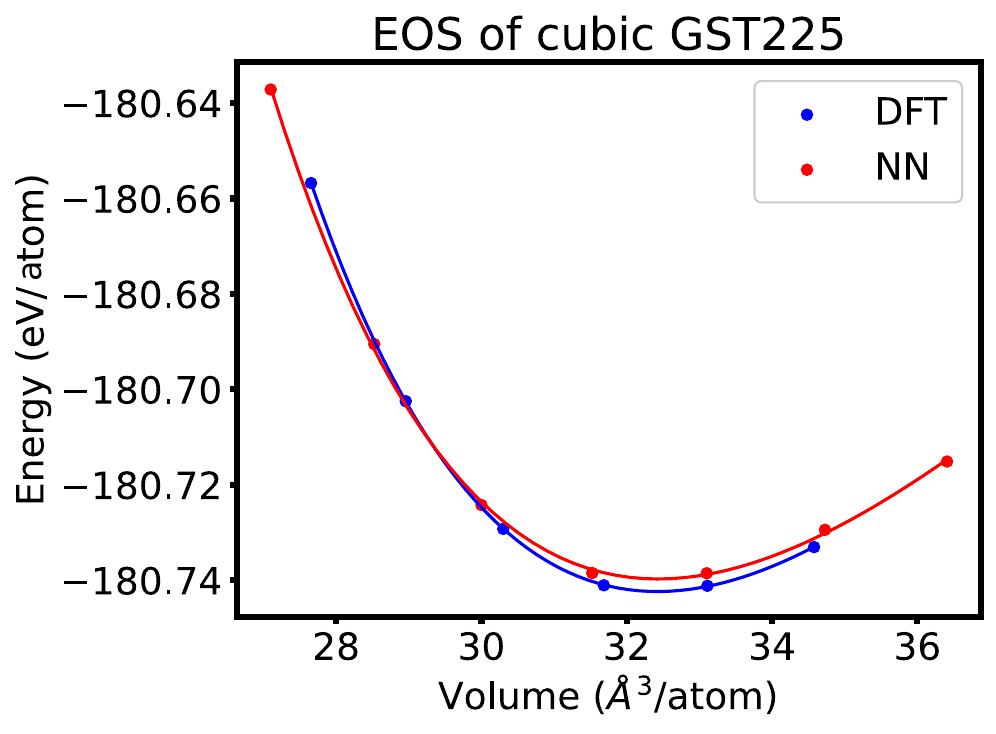}
 \caption{Equation of state (EOS) of the cubic phase of GST225 from DFT and NN calculations.}
 \label{fig:225EOScubic}
\end{figure}

\begin{figure}[H]
 \centering \includegraphics[width=0.6\textwidth, keepaspectratio]{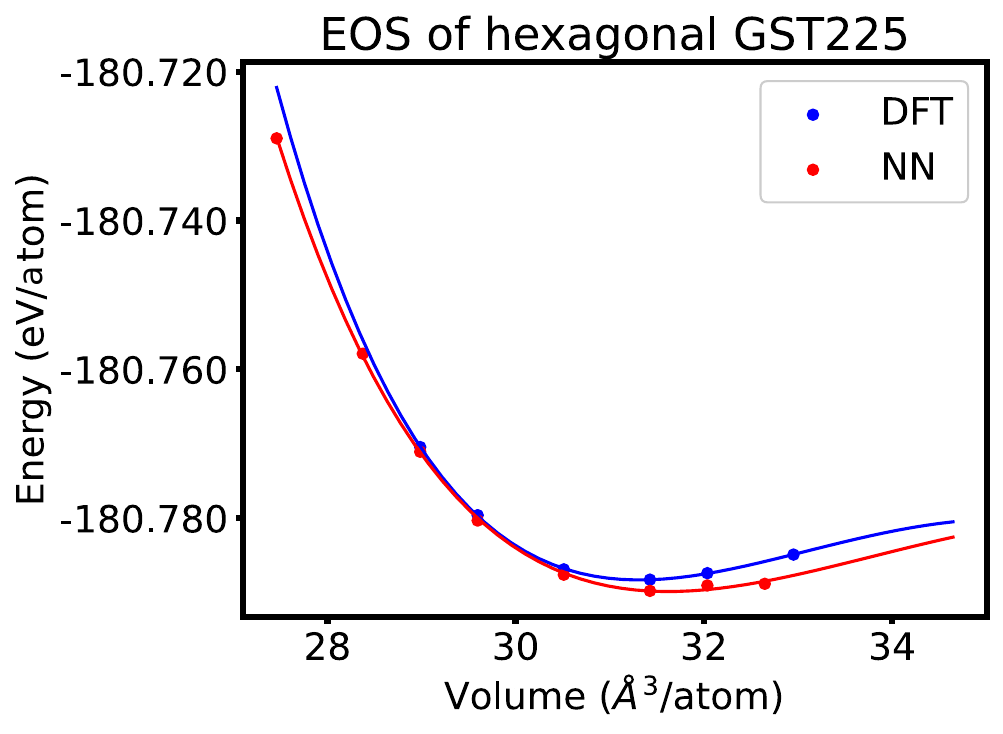}
 \caption{Equation of state (EOS) of the hexagonal phase of GST225 (Kooi stacking \cite{kooi}, see article)  from DFT and NN
calculations.}
 \label{fig:225EOSHex}
\end{figure}

The structural properties of the liquid Ge$_2$Sb$_2$Te$_5$ (GST225) at the experimental density of the amorphous phase  (0.0309 atoms/\AA$^3$)\cite{density225} obtained from Neural Network molecular dynamics (NN-MD) simulation with a 999-atom model at 990 K are compared with those obtained from DFT-MD simulation with a 216-atom model from Ref. \cite{npjOmar} at the same density (see Fig. \ref{fig:225liquid}).  The structural properties were averaged over trajectories 50 ps long.  The pair correlation functions, the bond angles distribution function and the distribution of coordination numbers are compared with DFT results in Fig. \ref{fig:225liquid}. The position of the first maximum and minimum of each partial correlation function are given in Table \ref{tab:gdrmax225liquid}. We also validated the NN potential over the dynamical properties of the liquid/supercooled phase down to 650 K, by comparing the self diffusion coefficient (D) from NN-MD simulation with DFT results from Ref. \cite{caravatiDIFF} (See Fig. \ref{fig:225liquid}d).  Then we validated the NN potential on the structural properties of the amorphous phase at the same density. The 999-atom liquid model was quenched from the melt at 990 K to 300 K in 100 ps, and equilibrated at 300 K for 40 ps. Structural properties were averaged over a subsequent run 50 ps long. In Fig. \ref{fig:225amorphous}, structural properties from NN-MD are compared with those obtained from amorphous model generated with DFT-MD simulation in Ref. \cite{npjOmar} with a similar protocol. The pair correlation functions, the bond angle distribution function, the distribution of coordination numbers and the distribution of the local order parameter $q$ for tetrahedricity for four-coordinated Ge atom are compared to DFT results in Fig. \ref{fig:225amorphous}. The position of the first maximum and minimum of each partial correlation function are given in Table \ref{tab:agdrmax}, while the average coordination number for different pairs of atoms are given in Table \ref{tab:am225NN}. In the amorphous phase of GST225, four-coordinated Ge atoms are in defective octahedral geometry and in tetrahedral configurations. The fraction of Ge atoms tetrahedral configuration can be quantified by integrating the $q$ order parameter between 0.8 and 1 as was discussed in previous works \cite{npjOmar}. In the NN model, 22\% of Ge atoms are in tetrahedral configuration, to be compared with 23\% from DFT models reported in Ref. \cite{npjOmar}. Finally, the NN potential was validated on the structural properties of the  hexagonal and cubic crystalline phases of GST225  by fitting the equation of state (EOS) at zero temperature  of both phases by fitting the energy vs volume data with the Birch–Murnaghan formula (see  Figs. \ref{fig:225EOScubic} and \ref{fig:225EOSHex}). The resulting parameters at equilibrium are compared with DFT data in Tables \ref{tab:225pEOScubic} and \ref{tab:225pEOShex}. Overall, the structural properties obtained from NN-MD simulations are in excellent agreement with those obtained from DFT-MD,  which suggests that the NN potential reliably describes the properties of the different phases of GST225. 

\newpage

\begin{center}
\captionof{table}{First maximum and minimum of the pair correlation functions in liquid GST225 at 990 K from NN-MD and DFT-MD (in parenthesis) simulations. Details on the simulation cells  are given in the caption of Fig. \ref{fig:225liquid}.}
\label{tab:gdrmax225liquid}

\begin{tabular}{c c c}
\hline
\hline
Bond type & First maximum (\AA) & First minimum (\AA) \\
\hline
GeGe& 2.71 (2.70)&3.29 (3.24)\\
GeSb& 2.86 (2.88)&3.51 (3.6)\\
GeTe& 2.81 (2.80)&-\\
SbSb&3.04 (3.00)&3.72 (3.78)\\
SbTe&3.00 (3.04)&-\\
TeTe&-&-\\
\hline
\hline
\end{tabular}
\end{center}

\begin{center}
\captionof{table}{Position (\AA) of the first maximum and minimum of the pair correlation functions in the amorphous phase of GST225 at 300 K from DFT and NN simulations. DFT data are given in parenthesis.}
\begin{tabular}{c c c}
\hline
\hline
Bond type & First maximum (\AA) & First minimum (\AA) \\
\hline
GeGe&2.64 (2.66)&3.14(3.15)\\
GeSb&2.82 (2.83)&3.3 (3.3)\\
GeTe&2.78 (2.78)&-\\
SbSb&2.92 (2.99)&-\\
SbTe&2.92 (2.97)&-\\
TeTe&3.00 (2.99)&-\\
\hline
\hline
\end{tabular}
\label{tab:agdrmax}

\end{center}{}

\begin{center}
\captionsetup{type=Table}
\captionof{table}{Average coordination numbers for different pairs of atoms in amorphous GST225 at 300 K generated from  DFT (in parenthesis) and NN simulations. Error bars for DFT data are obtained  from the analysis of four models. Details on the simulation cells and bonding cutoff  are given in the caption of Fig. \ref{fig:225amorphous}.}
\label{tab:am225NN}

\begin{tabular}{ c c c c c c  }
\hline

-       & Ge &          Sb & Te \\ \hline
Total   & 4.16 (4.17$\pm$0.1) &4.29 (4.40$\pm$0.05)&3.17 (3.16$\pm$0.05)\\
With Ge &0.28  (0.34$\pm$0.05)&0.34 (0.26$\pm$0.05)&1.42 (1.42$\pm$0.02)\\
With Sb & 0.34 (0.26$\pm$0.05) &0.41 (0.51$\pm$0.1) &1.42 (1.45$\pm$0.05)\\
With Te & 3.54 (3.57$\pm$0.05)&3.54 (3.63$\pm$0.1) &0.34 (0.28$\pm$0.02)\\
\hline
\\
\end{tabular}
\end{center}

\begin{center}
\captionsetup{type=Table}
\captionof{table}{Fitting parameters of the Birch-Murnaghan equation of state of cubic GST225  from NN and DFT-PBE calculations at zero temperature. The energy ($E_0$), volume ($V_0$), bulk modulus ($B$) and derivative of $B$ ($B^\prime$) at equilibrium are reported.}
\label{tab:225pEOScubic}
\begin{tabular}{c c c c c}

\hline
 & $E_0$ (eV atom$^{-1}$)& $V_0$ (\AA$^3$ atom$^{-1}$)& $B$ (GPa) & $B^\prime$ \\
\hline
DFT  & -180.7424 & 32.42&25.1&7.53\\
NN  & -180.7400 & 32.34&22.88&7.49\\
\hline
\\
\end{tabular}

\end{center}{}

\begin{center}
\captionsetup{type=Table}
\captionof{table}{Fitting parameters of the Birch-Murnaghan equation of state of hexagonal GST225  from NN and DFT-PBE calculations at zero temperature. The energy ($E_0$), volume ($V_0$), bulk modulus ($B$) and derivative of $B$ ($B^\prime$) at equilibrium are reported.}
\label{tab:225pEOShex}
\begin{tabular}{c c c c c}

\hline
 & $E_0$ (eV atom$^{-1}$)& $V_0$ (\AA$^3$ atom$^{-1}$)& $B$ (GPa) & $B^\prime$ \\
\hline
DFT  & -180.7883 & 31.31 &19.8&21.75\\
NN  & -180.7890 & 31.61 &16.5&18.5\\
\hline
\\
\end{tabular}

\end{center}{}

\subsection{$Ge_5Sb_2Te_3$}
\begin{figure}[H]
    \centering
    \includegraphics[width=0.8\linewidth,keepaspectratio]{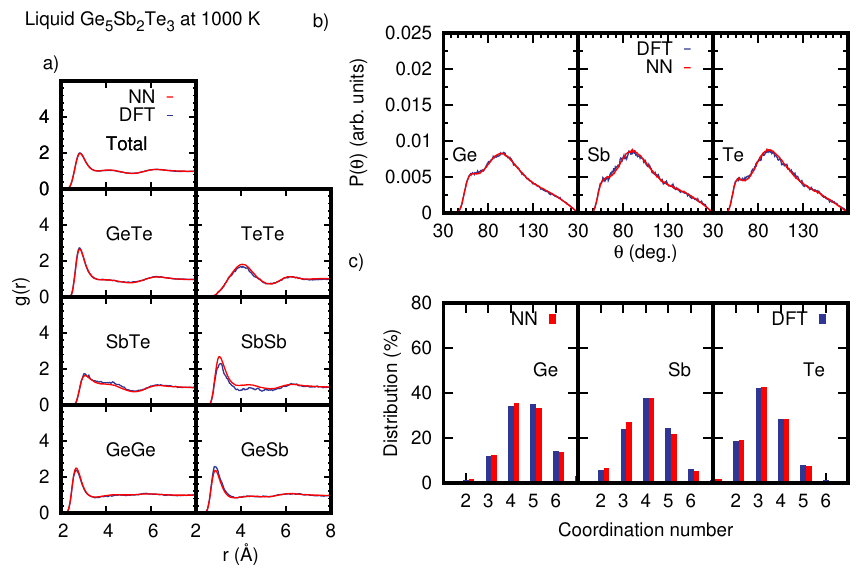}
    \caption{Structural properties of liquid GST523 at 1000 K from NN (red curves) and DFT (blue curves) simulations. NN (DFT) data refer to a 1760-atom (220-atom) model. a) Total and partial pair correlation functions. b) Angular distribution function resolved per central atomic species. The data are normalized to the number of triplets in each model. c) Distribution of coordination numbers resolved per chemical species, obtained by integrating the pair correlation functions up to a bonding cutoff corresponding to 3.2 \AA\ for all pairs except for Sb-Te for which a longer cutoff value of 3.4 \AA\ was used which correspond to the same cutoff used for the amorphous phase.}
    \label{fig:523liquid}
\end{figure}

\begin{figure}[H]
    \centering
    \includegraphics[width=0.8\linewidth,keepaspectratio]{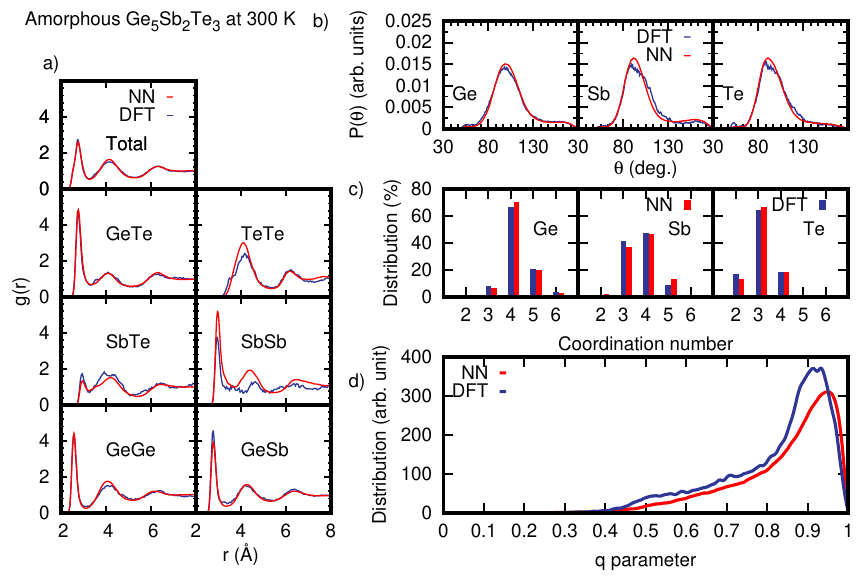}
    \caption{Structural properties of amorphous GST523 at 300 K from NN (red curves) and DFT (blue curves) simulations. NN (DFT) data refer to a 1760-atom (220-atom) model. a) Total and partial radial distribution functions. b) Angular distribution function resolved per central atomic species. The data are normalized to the number of triplets in each model. c) Distribution of coordination numbers resolved per chemical species, obtained by integrating the pair correlation functions up to a bonding cutoff corresponding to 3.2 \AA\ for all pairs except for Sb-Te for which a longer cutoff value of 3.4 \AA\ was used.  d) Distribution of the local order parameter $q$ for tetrahedricity for four-coordinated Ge atoms for the NN model and for each of our four DFT models. The large misfit in the Sb-Sb pair correlation function is due to the poor statistics in the small 220-atom cell for this composition with a low fraction of Sb. }
    \label{fig:523amorphous}
\end{figure}

As no experimental data are available for the density of amorphous $Ge_5Sb_2Te_3$ (GST523),  we have modeled GST523 at the theoretical density of 0.0338 atom/\AA\ obtained from DFT+D3 MD simulations in Ref. \cite{Omar2021}. However, in this section, we first report on structural properties from NN-MD and DFT-MD with no vdW corrections for the sake of validating  our NN potential fitted on a DFT-PBE database. The structural properties of the liquid GST523 at the theoretical density of amorphous phase obtained from NN-MD simulation with a 1760-atom model at 1000 K are compared to those obtained from DFT-MD simulation with a 220-atom model at the same density (see Fig. \ref{fig:523liquid}).  The structural properties were averaged over a trajectory  50 ps long.  The pair correlation functions, the bond angles distribution function and the distribution of coordination numbers are compared with DFT results in Fig. \ref{fig:523liquid}. Then, we validated the NN potential on the structural properties of the amorphous phase at the same density. The 1760-atom liquid model was quenched from the melt from 1000 K to 300 K in 100 ps, and then equilibrated at 300 K for 40 ps.
The structural properties were averaged over a subsequent run  50 ps long. In Fig. \ref{fig:523amorphous}, the structural properties from NN-MD are compared with those obtained from the amorphous model generated by DFT-MD with a similar protocol. The pair correlation functions, the bond angles distribution function, the distribution of coordination numbers and the distribution of the local order parameter $q$ for tetrahedricity for four-coordinated Ge atom are compared with DFT results in Fig. \ref{fig:523amorphous}. The average coordination number for different pairs of atoms are given in Table \ref{tab:am523NN}. We also quantified the fraction of four-coordinated Ge atoms in tetrahedral geometry by integrating the distribution of  the $q$ parameter between 0.8 and 1 as discussed  previously, resulting in 38\% in the NN model compared to 44\% in the DFT model. Once again, the structural properties obtained from NN-MD simulations are in excellent agreement with those obtained from DFT-MD which suggests that the NN potential reliably describes the properties of the different phases of GST523.

\newpage

\begin{center}
\captionsetup{type=Table}
\captionof{table}{Average coordination numbers for different pairs of atoms in amorphous GST523 at 300 K generated from  DFT (in parenthesis) and NN simulations. Details on the simulation cells and bonding cutoff  are given in the caption of Fig. \ref{fig:523amorphous}}
\label{tab:am523NN}

\begin{tabular}{ c c c c c c  }
\hline

-       & Ge          &        Sb  & Te \\ \hline
Total   & 4.18 (4.20) &3.73 (3.68) &3.08 (3.03)\\
With Ge & 1.86 (1.86) &1.92 (2.11) &2.59 (2.49)\\
With Sb & 0.77 (0.84) &1.08 (0.77) &0.48 (0.53)\\
With Te & 1.55 (1.49) &0.72 (0.79) &0.001 (0.005)\\
\hline
\\
\end{tabular}
\end{center}

\begin{figure}[H]
    \centering
    \includegraphics[width=0.8\linewidth,keepaspectratio]{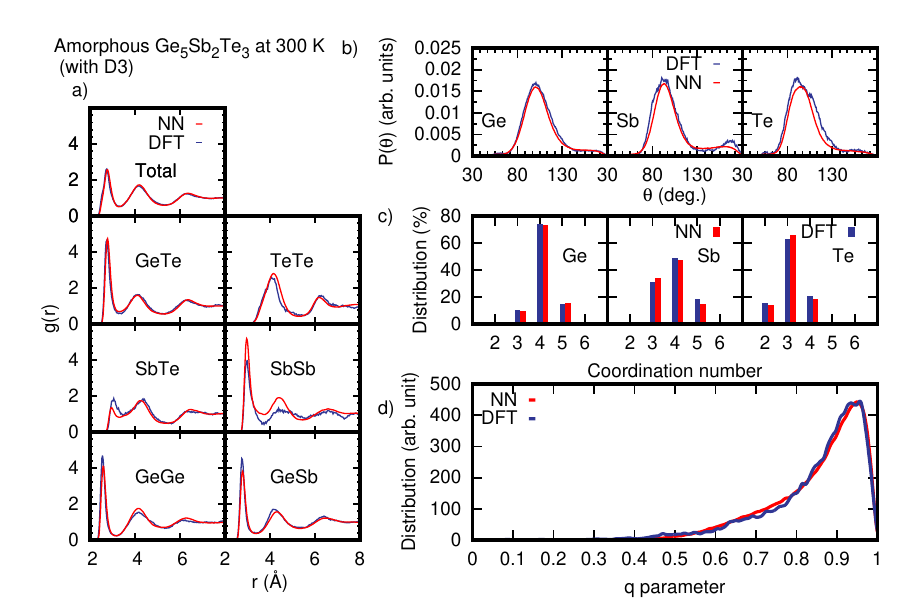}
    \caption{Structural properties of amorphous GST523 at 300 K from NN+D3 (red curves) and DFT (blue curves) simulations. NN (DFT) data refer to a 1760-atom (220-atom) model.  a) Total and partial radial distribution functions. b) Angular distribution function resolved per central atomic species. The data are normalized to the number of triplets in each model. c) Distribution of coordination numbers resolved per chemical species, obtained by integrating the pair correlation functions up to a bonding cutoff corresponding to 3.2 \AA\ for all pairs except for Sb-Te for which a longer cutoff value of 3.4 \AA\ was used.  d) Distribution of the local order parameter $q$ for tetrahedricity for four-coordinated Ge atoms.}
    \label{fig:523amorphousD3}
\end{figure}

We also computed the structural properties by including the vdW interaction D3 for the amorphous phase as discussed  above. The NN+D3 data are compared to the DFT+D3 results from Ref. \cite{Omar2021} in Fig. \ref{fig:523amorphousD3}

\subsection{$GeTe$}

\begin{figure}[H]
    \centering
    \includegraphics[width=0.8\linewidth,keepaspectratio]{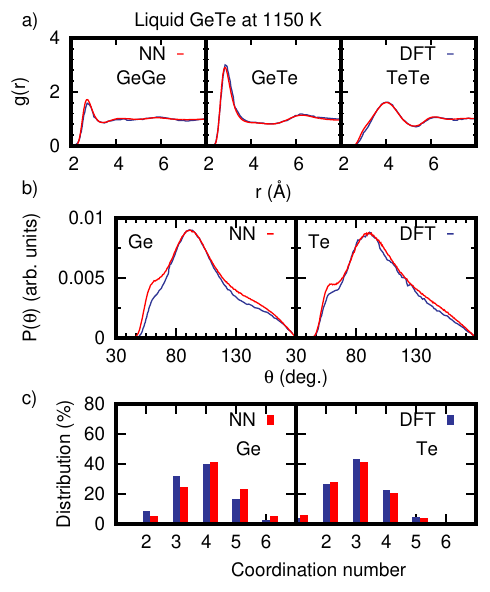}
    \caption{Structural properties of liquid GeTe at 1150 K from NN (red curves) and DFT (blue curves) simulations. NN (DFT) data refer  to a 4096-atom (300-atom) model. a) Total and partial radial distribution functions. b) Angular distribution function resolved per central atomic species. The data are normalized to the number of triplets in each model. c) Distribution of coordination numbers resolved per chemical species, obtained by integrating the pair correlation functions up to a bonding cutoff of 3.0 \AA\ for Ge-Ge and Te-Te and  of 3.22 \AA\  for Ge-Te. of the  We used the same cutoff values of the amorphous phase.}
    \label{fig:GeTeliquid}
\end{figure}

\begin{figure}[H]
    \centering
    \includegraphics[width=0.7\linewidth,keepaspectratio]{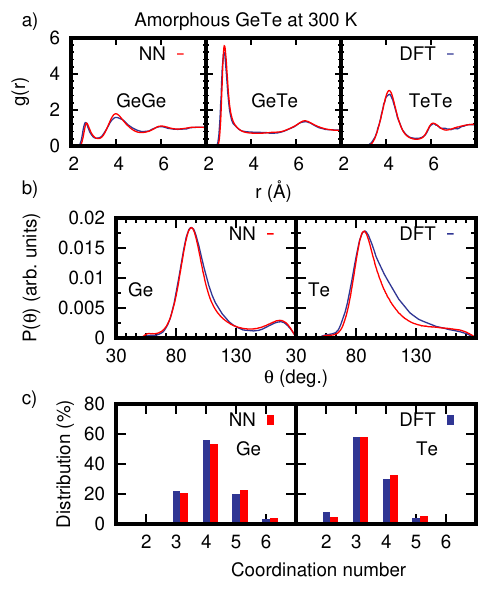}
    \includegraphics[width=0.7\linewidth,keepaspectratio]{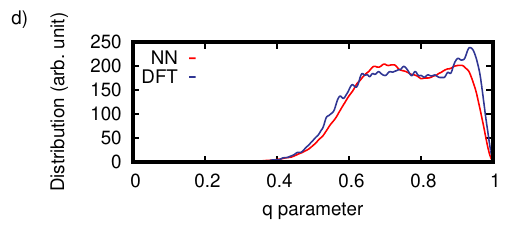}
    \caption{Structural properties of amorphous GeTe at 300 K from NN (red curves) and DFT (blue curves) simulations. NN (DFT) data refer  to a 4096-atom (300-atom) model. a) Total and partial radial distribution functions. b) Angular distribution function resolved per central atomic species. The data are normalized to the number of triplets in each model. c) Distribution of coordination numbers resolved per chemical species, obtained by integrating the pair correlation functions up to a bonding cutoff of 3.0 \AA\ for Ge-Ge and Te-Te and  of 3.22 \AA\  for Ge-Te. d) Distribution of the local order parameter $q$ for tetrahedricity for four-coordinated Ge atoms.}
    \label{fig:GeTeamorphous}
\end{figure}

\begin{figure}[H]
 \centering
 \includegraphics[width=0.6\textwidth, keepaspectratio]{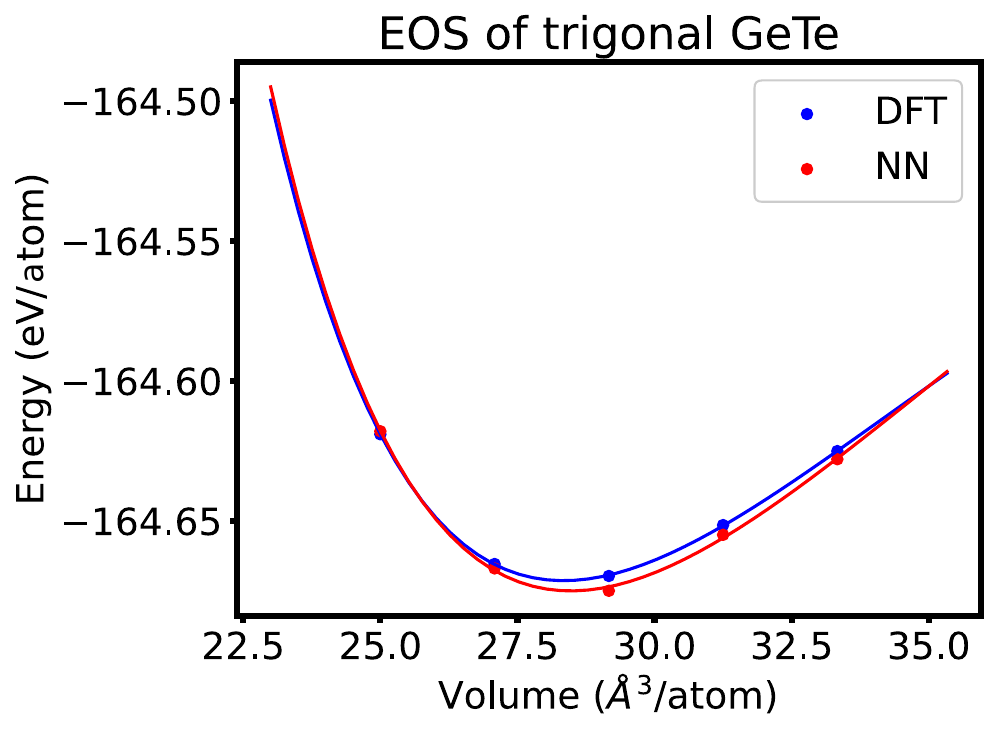}
 \caption{Equation of state (EOS) of the trigonal phase of GeTe from DFT and NN
calculations.}
 \label{fig:GeTeEOSHex}
\end{figure}

The structural properties of the liquid GeTe at the experimental density of amorphous phase  (0.0333 atoms/\AA$^3$)\cite{ghezzi} obtained from NN-MD simulation with a 4096-atom model at 1150 K are compared with those obtained from DFT-MD simulation with a 300-atom model from Ref. \cite{baratella2025} at the same density (see Fig. \ref{fig:GeTeliquid}).  The structural properties were averaged over a trajectory  50 ps long.  The pair correlation functions, the bond angles distribution function and the distribution of coordination numbers are compared with DFT results in Fig. \ref{fig:GeTeliquid}. Then, we validated the NN potential on the structural properties of the amorphous phase at the same density. The 4096-atom liquid model was quenched from the melt at 1150 K to 300 K in 100 ps, and equilibrated  for 40 ps at 300 K. Structural properties are averaged over a subsequent run 50 ps long. In Fig. \ref{fig:GeTeamorphous}, the structural properties from NN-MD are compared with those obtained from an amorphous model generated by DFT-MD simulation in Ref. \cite{baratella2025} with a similar protocol. The pair correlation functions, the bond angles distribution function, the distribution of coordination numbers and the distribution of the local order parameter $q$ for tetrahedricity for four-coordinated Ge atom are compared with DFT results in Fig. \ref{fig:GeTeamorphous}. The average coordination number for different pairs of atoms are given in Table \ref{tab:amGeTeNN}. We also quantified the fraction of four-coordinated Ge atoms in tetrahedral geometry by integrating the distribution  the $q$ parameter between 0.8 and 1 as discussed in previously, resulting in 21\% in NN model compared to 24\% in DFT results from Ref. \cite{baratella2025}. Finally, the NN potential was validated on the structural properties of the trigonal crystal phase of GeTe by fitting the EOS (see  Figs. \ref{fig:GeTeEOSHex}). The resulting parameters at equilibrium are compared with DFT data in Tables \ref{tab:GeTepEOShex}. Similarly to GST225 and GST523, the structural properties of GeTe obtained from NN-MD simulations are in a  good agreement with those obtained from DFT-MD, so we conclude that the NN potential can reliably describe the properties of the different phases of GeTe.

\begin{center}
\captionsetup{type=Table}
\captionof{table}{Average coordination numbers for different pairs of atoms in amorphous $GeTe$ at 300 K generated from  DFT (in parenthesis) and NN simulations. Details on the simulation cells and bonding cutoff are given in the caption of Fig. \ref{fig:GeTeamorphous}.}
\label{tab:amGeTeNN}

\begin{tabular}{ c c c c c  }
\hline

-        &        Ge  & Te \\ \hline
Total    &4.09 (4.04) &3.38 (3.21)\\
With Ge  &0.71 (0.87) &3.37 (3.18)\\
With Te  &3.37 (3.18) &0.01 (0.04)\\
\hline
\\
\end{tabular}
\end{center}

\begin{center}
\captionsetup{type=Table}
\captionof{table}{Fitting parameters of the Birch-Murnaghan equation of state of hexagonal GeTe  from NN and DFT-PBE calculations at zero temperature. The energy ($E_0$), volume ($V_0$), bulk modulus ($B$) and derivative of $B$ ($B^\prime$) at equilibrium are reported.}
\label{tab:GeTepEOShex}
\begin{tabular}{c c c c c}

\hline
 & $E_0$ (eV atom$^{-1}$)& $V_0$ (\AA$^3$ atom$^{-1}$)& $B$ (GPa) & $B^\prime$ \\
\hline
DFT & -164.671 &28.34 &28.88 &8.55\\
NN  & -164.675 &28.47 &29.61 &7.93\\

\hline
\\
\end{tabular}

\end{center}{}

\subsection{$Ge_2Te$}
\begin{figure}[H]
    \centering
    \includegraphics[width=0.8\linewidth,keepaspectratio]{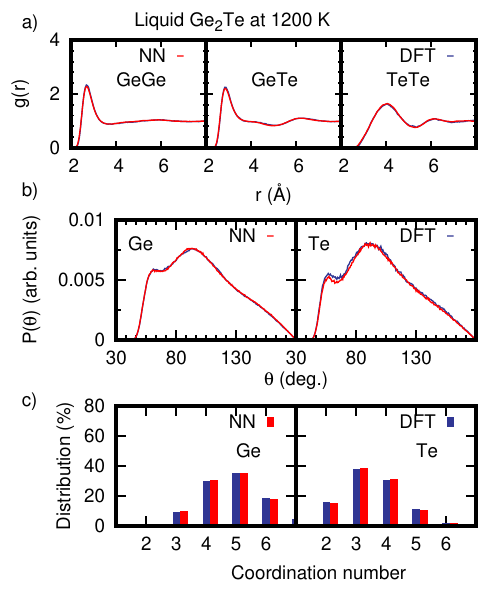}
    \caption{Structural properties of liquid Ge$_2$Te at 1200 K from NN (red curves) and DFT (blue curves) simulations. NN (DFT) data refer  to a 2400-atom (300-atom) model. a) Total and partial radial distribution functions. b) Angular distribution function resolved per central atomic species. The data are normalized to the number of triplets in each model. c) Distribution of coordination numbers resolved per chemical species, obtained by integrating the pair correlation functions up to a bonding cutoff of 3.0 \AA\ for Ge-Ge and Te-Te and  of 3.22 \AA\  for Ge-Te. We used the same cutoff values of the amorphous phase.}
    \label{fig:Ge2Teliquid}
\end{figure}

\begin{figure}[H]
    \centering
    \includegraphics[width=0.7\linewidth,keepaspectratio]{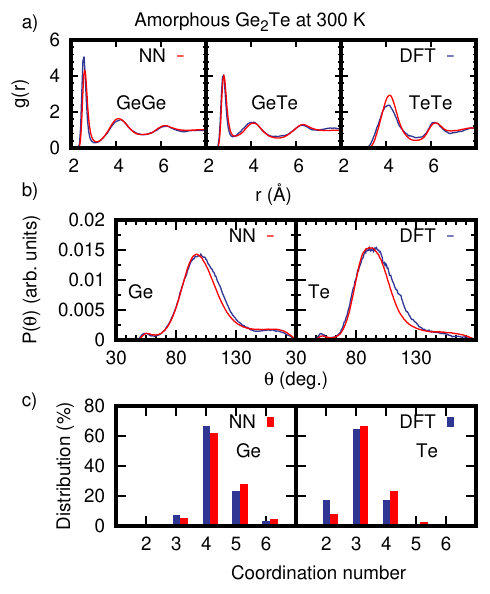}
    \includegraphics[width=0.7\linewidth,keepaspectratio]{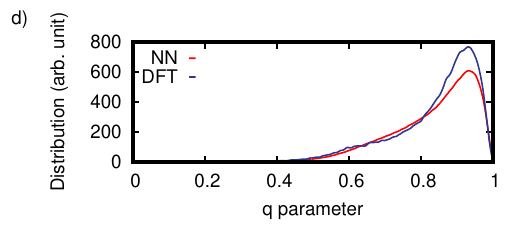}
    \caption{Structural properties of amorphous Ge$_2$Te at 300 K from NN (red curves) and DFT (blue curves) simulations. NN (DFT) data refer  to a 2400-atom (300-atom) model. a) Total and partial radial distribution functions. b) Angular distribution function resolved per central atomic species. The data are normalized to the number of triplets in each model. c) Distribution of coordination numbers resolved per chemical species, obtained by integrating the pair correlation functions up to a bonding cutoff of 3.0 \AA\ for Ge-Ge and Te-Te and  of 3.22 \AA\  for Ge-Te. d) Distribution of the local order parameter $q$ for tetrahedricity for four-coordinated Ge atoms. }
    \label{fig:Ge2Teamorphous}
\end{figure}

Similarly to GST523, in the lack of experimental data for the density of Ge$_2$Te, we have modeled this alloy at the theoretical density of 0.0355 atom/\AA$^3$ obtained from DFT+D3 MD simulations in Ref. \cite{baratella2025}. Yet, in this section we report on structural properties from  NN-MD and DFT-MD with no vdW corrections for the sake of validating the fitting the NN potential fitted on a DFT-PBE database.
The structural properties of the liquid Ge$_2$Te at the theoretical density of amorphous phase  (0.0355 atoms/\AA$^3$) obtained from NN-MD simulation with a 2400-atom model at 1200 K are compared with those obtained from DFT-MD simulation with a 300-atom model from Ref. \cite{baratella2025} at the same density (see Fig. \ref{fig:Ge2Teliquid}).  The structural properties were averaged over a trajectory  50 ps long.  The pair correlation functions, the bond angles distribution function and the distribution of coordination numbers are compared with DFT results in Fig. \ref{fig:Ge2Teliquid}. Then, we validated the NN potential on the structural properties of the amorphous phase at the same density. The 4096-atom liquid model was quenched from the melt at 1200 K to 300 K in 100 ps, and equilibrated at 300 K for 40 ps. Structural properties averaged over a subsequent run 50 ps long. In Fig. \ref{fig:Ge2Teamorphous}, the structural properties from NN-MD are compared with those obtained from an amorphous model generated by DFT-MD simulation in Ref. \cite{baratella2025} with similar protocol. The pair correlation functions, the bond angles distribution function, the distribution of coordination numbers and the distribution of the local order parameter $q$ for tetrahedricity for four-coordinated Ge atom are compared with DFT results in Fig. \ref{fig:Ge2Teamorphous}. The average coordination number for different pairs of atoms are given in Table \ref{tab:amGe2TeNN}. We also quantified the fraction of four-coordinated Ge atoms in tetrahedral geometry by integrating the distribution of  the $q$ parameter between 0.8 and 1 as discussed  previously, resulting in 43\% in NN model compared to 52\% in DFT results from Ref. \cite{baratella2025}. Again, the structural properties of Ge$_2$Te obtained from NN-MD simulations are in a good agreement with those obtained from DFT-MD across the different phases of this alloy.

\begin{center}
\captionsetup{type=Table}
\captionof{table}{Average coordination numbers for different pairs of atoms in amorphous $Ge_2Te$ at 300 K generated from  DFT (in parenthesis) and NN simulations. Details on the simulation cells and bonding cutoff are given in the caption of Fig. \ref{fig:Ge2Teamorphous}.}
\label{tab:amGe2TeNN}

\begin{tabular}{ c c c c c  }
\hline

-        &        Ge  & Te \\ \hline
Total    &4.33 (4.04) &3.20 (2.96)\\
With Ge  &2.75 (2.58) &3.19 (2.96)\\
With Te  &1.57 (1.46) &0.01 (0.00)\\
\hline
\\
\end{tabular}
\end{center}

\subsection{$GeSb$}

In the lack of experimental data for the density of GeSb,  we have modeled this alloy at the theoretical density of 0.0352 atom/\AA$^3$. This density was obtained from a NPT simulation with the NN potential at 1000 K. In this simulation, we first started with a density of 0.0368 atom/\AA$^3$, 8\% lower than the density of the crystal (0.04 atom/\AA$^3$) as was suggested for GST225. After an  equilibration MD simulation 30 ps  long, we obtained the theoretical density as the average over 30 ps long trajectory which results in average density 0.0352 atom/\AA$^3$ which is slightly lower than our initial estimate. We then verified that the residual stress at this density in the liquid at 1000 K is small in a DFT simulation 20 ps long, which is on average 0.89 GPa.

\begin{figure}[H]
    \centering
    \includegraphics[width=0.8\linewidth,keepaspectratio]{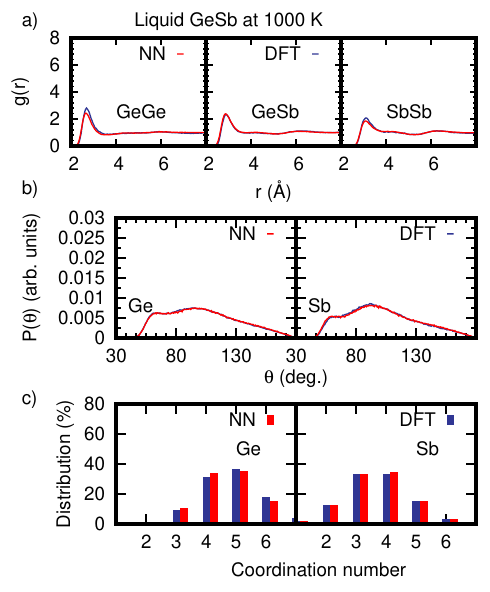}
    \caption{Structural properties of liquid GeSb at 1000 K from NN (red curves) and DFT (blue curves) simulations. NN (DFT) data refer  to a 512-atom (148-atom) model. a) Total and partial radial distribution functions. b) Angular distribution function resolved per central atomic species. The data are normalized to the number of triplets in each model. c) Distribution of coordination numbers resolved per chemical species, obtained by integrating the pair correlation functions up to a bonding cutoff of 3.2 \AA\ for all pairs. We used the same bonding cutoff of the amorphous phase.}
    \label{fig:GeSbliquid}
\end{figure}

\begin{figure}[H]
    \centering
    \includegraphics[width=0.7\linewidth,keepaspectratio]{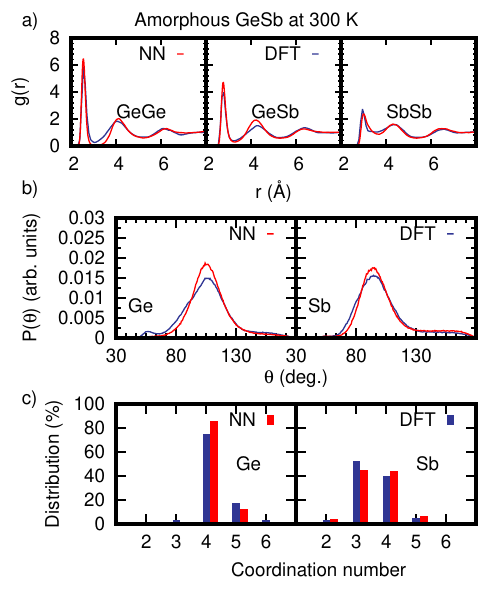}
    \includegraphics[width=0.7\linewidth,keepaspectratio]{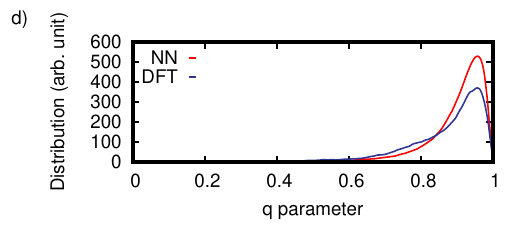}
    \caption{Structural properties of amorphous GeSb at 300 K from NN (red curves) and DFT (blue curves) simulations. NN (DFT) data refer  to a 512-atom (148-atom) model. a) Total and partial radial distribution functions. b) Angular distribution function resolved per central atomic species. The data are normalized to the number of triplets in each model. c) Distribution of coordination numbers resolved per chemical species, obtained by integrating the pair correlation functions up to a bonding cutoff of 3.2 \AA\ for all pairs. d) Distribution of the local order parameter $q$ for tetrahedricity for four-coordinated Ge atoms.}
    \label{fig:GeSbamorphous}
\end{figure}

The structural properties of the liquid GeSb at the theoretical density of the liquid phase  (0.0352 atoms/\AA$^3$) obtained from NN-MD simulation with a 512-atom model at 1000 K are compared with those obtained from DFT-MD simulation with a 148-atom model at the same density (see Fig. \ref{fig:GeSbliquid}).  The structural properties were averaged over a trajectory  50 ps long.  The pair correlation functions, the bond angles distribution function and the distribution of coordination numbers are compared with DFT results in Fig. \ref{fig:GeSbliquid}. Then, we validated the NN potential on the structural properties of the amorphous phase at the same density. The 512-atom liquid model was quenched from the melt at 1000 K to 300 K in 100 ps, and equilibrated at 300 K for  40 ps. Structural properties are averaged over a subsequent run 50 ps long. In Fig. \ref{fig:GeSbamorphous}, the structural properties from NN-MD are compared with those obtained from amorphous model generated with DFT-MD with similar protocol. The pair correlation functions, the bond angles distribution function, the distribution of coordination numbers and the distribution of the local order parameter $q$ for tetrahedricity for four-coordinated Ge atom are compared with DFT results in Fig. \ref{fig:GeSbamorphous}. The average coordination number for different pairs of atoms are given in Table \ref{tab:amGeSbNN}. We also quantified the fraction of four-coordinated Ge atoms in tetrahedral geometry by integrating the distribution of  the $q$ parameter between 0.8 and 1 as discussed  previously, resulting in 75\% in the NN model to be compared to the value of 63\% in the DFT model.

\begin{center}
\captionsetup{type=Table}
\captionof{table}{Average coordination numbers for different pairs of atoms in amorphous $GeSb$ at 300 K generated from  DFT (in parenthesis) and NN simulations.  Details on the simulation cells and bonding cutoff are given in the caption of Fig. \ref{fig:GeSbamorphous}.}
\label{tab:amGeSbNN}

\begin{tabular}{ c c c c c  }
\hline

-        &        Ge  & Sb \\ \hline
Total    &4.13 (4.24) &3.53 (3.47)\\
With Ge  &2.01 (2.27) &2.12 (1.97)\\
With Sb  &2.12 (1.97) &1.41 (1.50)\\
\hline
\\
\end{tabular}
\end{center}

\subsection{$Ge_{15}Sb_{85}$}

\begin{figure}[H]
    \centering
    \includegraphics[width=0.8\linewidth,keepaspectratio]{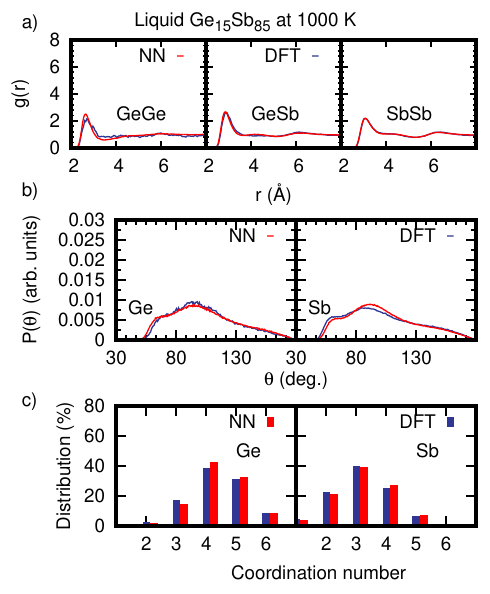}
    \caption{Structural properties of liquid Ge$_{15}$Sb$_{85}$ at 1000 K from NN (red curves) and DFT (blue curves) simulations. NN (DFT) data refer  to a 1000-atom (200-atom) model. a) Total and partial radial distribution functions. b) Angular distribution function resolved per central atomic species. The data are normalized to the number of triplets in each model. c) Distribution of coordination numbers resolved per chemical species, obtained by integrating the pair correlation functions up to a bonding cutoff of 2.9 \AA\ for  Ge-Ge, 3.1 \AA\ for Ge-Sb bonds and 3.5 \AA\ for Sb-Sb. We used the same bonding cutoff of the amorphous phase.}
    \label{fig:Ge15Sb85liquid}
\end{figure}

\begin{figure}[H]
    \centering
    \includegraphics[width=0.65\linewidth,keepaspectratio]{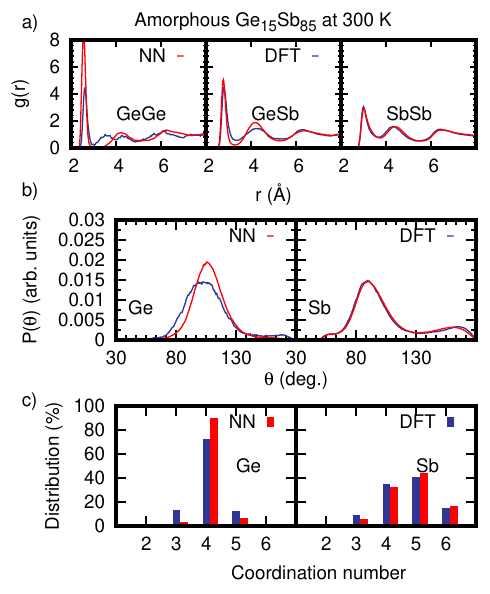}
    \includegraphics[width=0.7\linewidth,keepaspectratio]{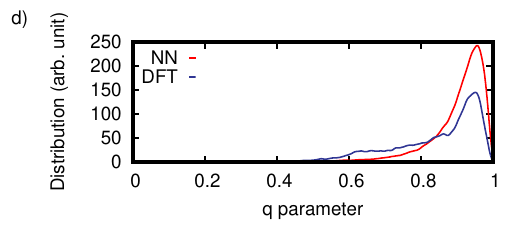}
    \caption{Structural properties of amorphous Ge$_{15}$Sb$_{85}$ at 300 K from NN (red curves) and DFT (blue curves) simulations. NN (DFT) data refer  to a 1000-atom (200-atom) model. a) Total and partial radial distribution functions. b) Angular distribution function resolved per central atomic species. The data are normalized to the number of triplets in each model. c) Distribution of coordination numbers resolved per chemical species, obtained by integrating the pair correlation functions up to a bonding cutoff of 2.9 \AA\ for  Ge-Ge, 3.1 \AA\ for Ge-Sb bonds and 3.5 \AA\ for Sb-Sb. d) Distribution of the local order parameter $q$ for tetrahedricity for four-coordinated Ge atoms.}
    \label{fig:Ge15Sb85amorphous}
\end{figure}

The structural properties of the liquid Ge$_{15}$Sb$_{85}$ at the experimental density of amorphous phase  (0.0321 atoms/\AA$^3$) \cite{zalden2010atomic} obtained from NN-MD simulation with a 1000-atom model at 1000 K are compared with those obtained from DFT-MD simulation with a 200-atom model at the same density (see Fig. \ref{fig:Ge15Sb85liquid}).  The structural properties were averaged over a trajectory  50 ps long.  The pair correlation functions, the bond angles distribution function and the distribution of coordination numbers are compared with DFT results in Fig. \ref{fig:Ge15Sb85liquid}. Then we validated the NN potential on the structural properties of the amorphous phase at the same density. The 1000-atom liquid model was quenched from the melt at 1000 K to 300 K in 100 ps, and equilibrated at 300 K for 40 ps. Structural properties are averaged over a subsequent run   50 ps long. In Fig. \ref{fig:Ge15Sb85amorphous}, the structural properties from NN-MD are compared with those obtained from amorphous model generated with DFT-MD with similar protocol. The pair correlation functions, the bond angles distribution function, the distribution of coordination numbers and the distribution of the local order parameter $q$ for tetrahedricity for four-coordinated Ge atom are compared with DFT results in Fig. \ref{fig:Ge15Sb85amorphous}. The average coordination number for different pairs of atoms are given in Table \ref{tab:amGe15Sb85NN}. We also quantified the fraction of four-coordinated Ge atoms in tetrahedral geometry by integrating the distribution of the $q$ parameter between 0.8 and 1 as discussed  previously, resulting in 81\% in the NN model to be compared to the value of 53\% in the DFT model. Similarly to other alloys, the structural properties of Ge$_{15}$Sb$_{85}$ obtained from NN-MD and DFT-MD simulations are in a  good agreement. The largest misfit is in the fraction of Ge atoms in tetrahedral configuration which is overestimated in NN data.  The large misfit in the Ge-Ge pair correlation function is due to the poor statistics in the small 200-atom cell for this composition with a low fraction of Ge.

\begin{center}
\captionsetup{type=Table}
\captionof{table}{Average coordination numbers for different pairs of atoms in amorphous $Ge_{15}Sb_{85}$ at 300 K generated from  DFT (in parenthesis) and NN simulations. Details on the simulation cells and bonding cutoff are given in the caption of Fig. \ref{fig:Ge15Sb85amorphous}.}
\label{tab:amGe15Sb85NN}

\begin{tabular}{ c c c c c  }
\hline

-        &        Ge  & Sb \\ \hline
Total    &4.03 (3.97) &4.75 (4.63)\\
With Ge  &0.73 (0.43) &0.58 (0.63)\\
With Sb  &3.30 (3.54) &4.17 (4.00)\\
\hline
\\
\end{tabular}
\end{center}

\subsection{$Sb_2Te_3$}
\begin{figure}[H]
    \centering
    \includegraphics[width=0.8\linewidth,keepaspectratio]{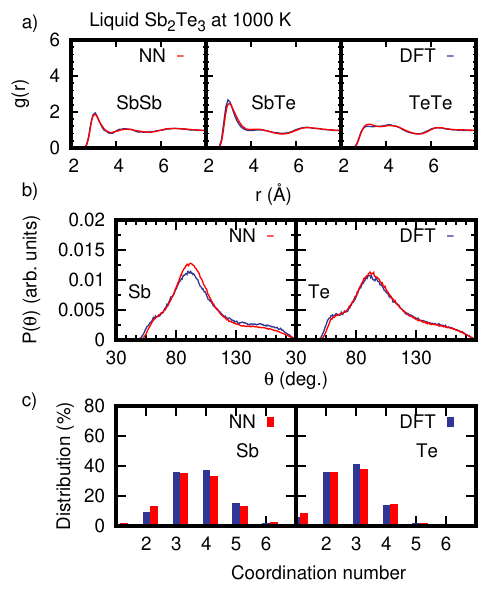}
    \caption{Structural properties of liquid Sb$_2$Te$_3$ at 1000 K from NN (red curves) and DFT (blue curves) simulations. NN (DFT) data refer  to a 960-atom (200-atom) model. a) Total and partial radial distribution functions. b) Angular distribution function resolved per central atomic species. The data are normalized to the number of triplets in each model. c) Distribution of coordination numbers resolved per chemical species, obtained by integrating the pair correlation functions up to a bonding cutoff of 3.2 \AA\ for  Sb-Sb and Te-Te and of 3.4 \AA\ for Sb-Te. We used the same bonding cutoff of the amorphous phase.}
    \label{fig:Sb2Te3liquid}
\end{figure}

\begin{figure}[H]
    \centering
    \includegraphics[width=0.8\linewidth,keepaspectratio]{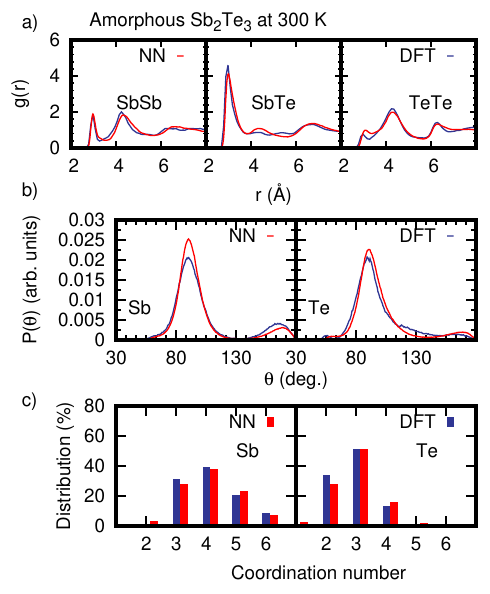}
    \caption{Structural properties of amorphous Sb$_2$Te$_3$ at 300 K from NN (red curves) and DFT (blue curves) simulations. NN (DFT) data refer  to a 960-atom (200-atom) model. a) Total and partial radial distribution functions. b) Angular distribution function resolved per central atomic species. The data are normalized to the number of triplets in each model. c) Distribution of coordination numbers resolved per chemical species, obtained by integrating the pair correlation functions up to a bonding cutoff of 3.2 \AA\ for Sb-Sb and Te-Te and of 3.4 \AA\ for Sb-Te.}
    \label{fig:Sb2Te3amorphous}
\end{figure}

\begin{figure}[H]
 \centering
 \includegraphics[width=0.6\textwidth, keepaspectratio]{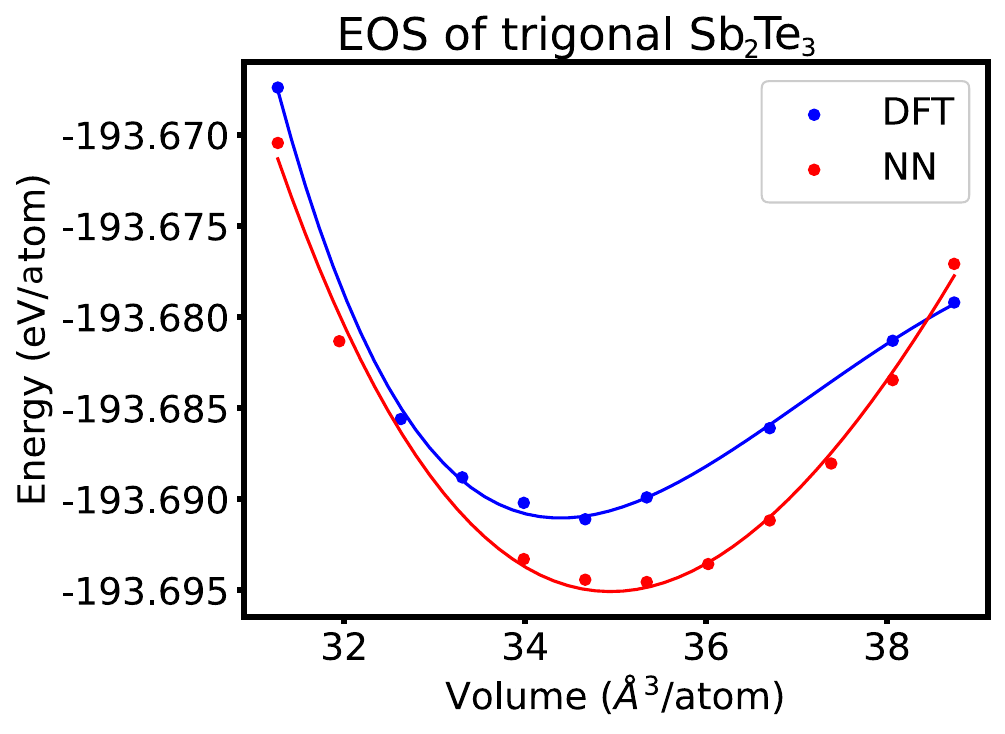}
 \caption{Equation of state (EOS) of the trigonal phase of Sb$_2$Te$_3$ from DFT and NN
calculations.}
 \label{fig:Sb2Te3EOSHex}
\end{figure}

The structural properties of  liquid Sb$_2$Te$_3$ at the experimental density of liquid phase  (0.0281 atoms/\AA$^3$ at 1003 K)\cite{singh2003structural} obtained from NN-MD simulation with a 960-atom model at 1000 K are compared with those obtained from DFT-MD simulation with a 200-atom model at the same density (see Fig. \ref{fig:Sb2Te3liquid}). The structural properties were averaged over a trajectory  50 ps long.  The pair correlation functions, the bond angles distribution function and the distribution of coordination numbers are compared with DFT results in Fig. \ref{fig:Sb2Te3liquid}. Then, we validated the NN potential on the structural properties of the amorphous phase at the same density. The 960-atom liquid model was quenched from the melt at 1000 K to 300 K in 100 ps, and equilibrated at 300 K  for 40 ps.  Structural properties  are averaged over a subsequent run 50 ps long. In Fig. \ref{fig:Sb2Te3amorphous}, the structural properties from NN-MD are compared with those obtained from amorphous model generated by DFT-MD simulation with a similar protocol. The pair correlation functions, the bond angles distribution function, and the distribution of coordination numbers in Fig. \ref{fig:Sb2Te3amorphous}. The average coordination number for different pairs of atoms are given in Table \ref{tab:amSb2Te3NN}. Finally, the NN potential was validated on the structural properties of the trigonal crystal phase of Sb$_2$Te$_3$ by fitting the EOS (see  Figs. \ref{fig:Sb2Te3EOSHex}). The resulting parameters at equilibrium are compared with DFT data in Tables \ref{tab:Sb2Te3pEOShex}. Overall, the structural properties of Sb$_2$Te$_3$ from NN-MD simulations are in a good agreement with those obtained from DFT-MD. The largest misfit is in the EOS of the trigonal phase with a 4 meV/atom error on the equilibrium  energy, which is still comparable to the average error of the NN potential.

\newpage

\begin{center}
\captionsetup{type=Table}
\captionof{table}{Average coordination numbers for different pairs of atoms in amorphous $Sb_2Te_3$ at 300 K generated from  DFT (in parenthesis) and NN simulations. Details on the simulation cells and bonding cutoff are given in the caption of Fig. \ref{fig:Sb2Te3amorphous}.}
\label{tab:amSb2Te3NN}

\begin{tabular}{ c c c c c  }
\hline

-        &        Sb  & Te \\ \hline
Total    &4.02 (4.05) &2.85 (2.80)\\
With Sb  &0.59 (0.56) &2.28 (2.32)\\
With Te  &3.42 (3.48) &0.57 (0.48)\\
\hline
\\
\end{tabular}
\end{center}

\begin{center}
\captionsetup{type=Table}
\captionof{table}{Fitting parameters of the Birch-Murnaghan equation of state of the trigonal phase of Sb$_2$Te$_3$ from NN and DFT-PBE calculations at zero temperature. The energy ($E_0$), volume ($V_0$), bulk modulus ($B$) and derivative of $B$ ($B^\prime$) at equilibrium are reported.}
\label{tab:Sb2Te3pEOShex}
\begin{tabular}{c c c c c}

\hline
 & $E_0$ (eV atom$^{-1}$)& $V_0$ (\AA$^3$ atom$^{-1}$)& $B$ (GPa) & $B^\prime$ \\
\hline
DFT  & -193.691& 34.40 &16.15&16.5\\
NN  & -193.695 & 34.94 &16.05&4.3\\
\hline
\\
\end{tabular}
\end{center}{}
\newpage
\subsection{$Sb_2Te$}

\begin{figure}[H]
    \centering
    \includegraphics[width=0.8\linewidth,keepaspectratio]{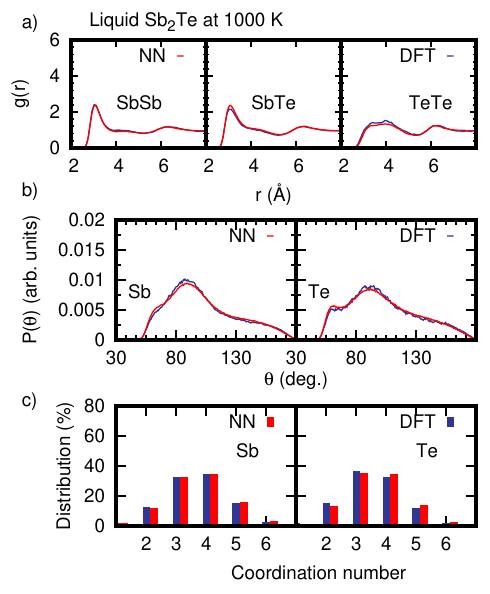}
    \caption{Structural properties of liquid $Sb_2Te$ at 1000 K from NN (red curves) and DFT (blue curves) simulations. NN (DFT) data refer  to a 1152-atom (198-atom) model. a) Total and partial radial distribution functions. b) Angular distribution function resolved per central atomic species. The data are normalized to the number of triplets in each model. c) Distribution of coordination numbers resolved per chemical species, obtained by integrating the pair correlation functions up to a bonding cutoff  of 3.2 \AA\ for  Sb-Sb and Te-Te and of 3.4 \AA\ for Sb-Te. We used the same bonding cutoff of the amorphous phase.}
    \label{fig:Sb2Teliquid}
\end{figure}

\begin{figure}[H]
    \centering
    \includegraphics[width=0.8\linewidth,keepaspectratio]{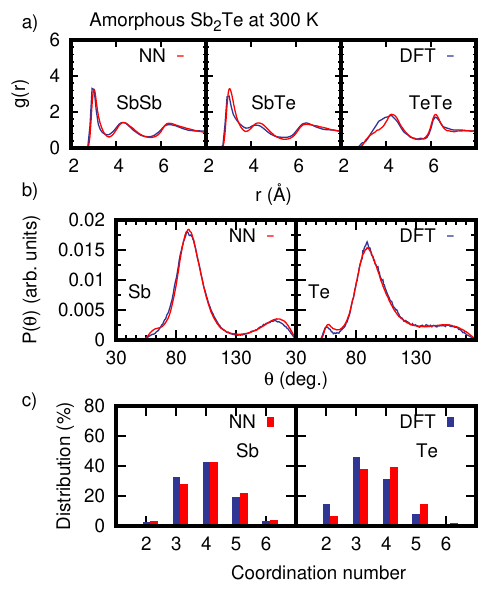}
    \caption{Structural properties of amorphous Sb$_2$Te at 300 K from NN (red curves) and DFT (blue curves) simulations. NN (DFT) data refer  to a 1152-atom (198-atom) model.  a) Total and partial radial distribution functions. b) Angular distribution function resolved per central atomic species. The data are normalized to the number of triplets in each model. c) Distribution of coordination numbers resolved per chemical species, obtained by integrating the pair correlation functions up to a bonding cutoff of 3.2 \AA\ for  Sb-Sb and Te-Te and of 3.4 \AA\ for Sb-Te.}
    \label{fig:Sb2Teamorphous}
\end{figure}

The structural properties of  liquid Sb$_2$Te at the experimental density of amorphous phase (0.0309 atoms/\AA$^3$)\cite{zheng2016candidate} obtained from NN-MD simulation with a 1152-atom model at 1000 K are compared with those obtained from DFT-MD simulation with a 198-atom model at the same density (see Fig. \ref{fig:Sb2Teliquid}). The structural properties were averaged over a trajectory  50 ps long.  The pair correlation functions, the bond angles distribution function and the distribution of coordination numbers are compared with DFT results in Fig. \ref{fig:Sb2Teliquid}. Then we validated the NN potential on the structural properties of the amorphous phase at the same density. The 1152-atom liquid model was quenched from the melt at 1000 K to 300 K in 100 ps,  and equilibrated at 300 for 40 ps.
Structural properties were averaged over a subsequent run  50 ps long. In Fig. \ref{fig:Sb2Teamorphous}, the structural properties from NN-MD are compared with those obtained from amorphous model generated with DFT-MD simulation with similar protocol. The pair correlation functions, the bond angles distribution function, and the distribution of coordination numbers in Fig. \ref{fig:Sb2Teamorphous}. The average coordination number for different pairs of atoms are given in Table \ref{tab:amSb2TeNN}.

\newpage
\begin{center}
\captionsetup{type=Table}
\captionof{table}{Average coordination numbers for different pairs of atoms in amorphous $Sb_2Te$ at 300 K generated from  DFT (in parenthesis) and NN simulations.  Details on the simulation cells and bonding cutoff are given in the caption of Fig. \ref{fig:Sb2Teamorphous}.}
\label{tab:amSb2TeNN}

\begin{tabular}{ c c c c c  }
\hline

-        &        Sb  & Te \\ \hline
Total    &3.95 (3.88) &3.67 (3.32)\\
With Sb  &2.17 (2.29) &3.55 (3.17)\\
With Te  &1.78 (1.59) &0.11 (0.15)\\
\hline
\\
\end{tabular}
\end{center}

\subsection{$Ge$}

\begin{figure}[H]
    \centering
    \includegraphics[width=0.4\linewidth,keepaspectratio]{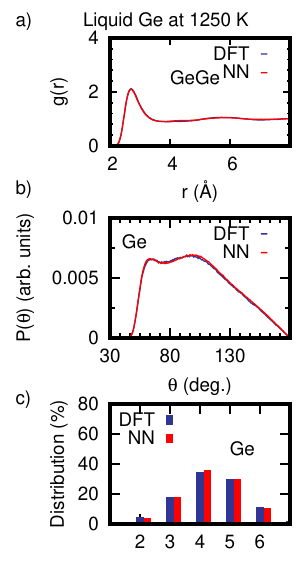}
    \caption{Structural properties of liquid Ge at 1250 K from NN (red curves) and DFT (blue curves) simulations. NN (DFT) data refer  to a 2400-atom (300-atom) model.  a) Total radial distribution functions. b) Angular distribution function resolved per central atomic species. The data are normalized to the number of triplets in each model. c) Distribution of coordination numbers resolved per chemical species, obtained by integrating the pair correlation functions up to a bonding cutoff of 3.0 \AA. We used the same bonding cutoff of the amorphous phase.}
    \label{fig:Geliquid}
\end{figure}

\begin{figure}[H]
    \centering
    \includegraphics[width=0.45\linewidth,keepaspectratio]{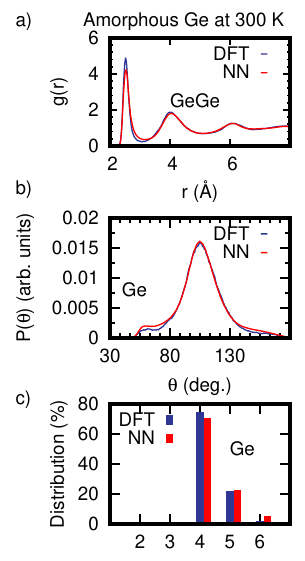}
    \includegraphics[width=0.55\linewidth,keepaspectratio]{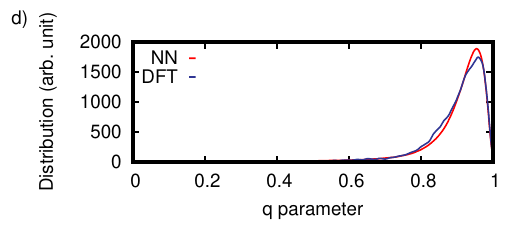}
    \caption{Structural properties of amorphous Ge at 300 K from NN (red curves) and DFT (blue curves) simulations. NN (DFT) data refer  to a 2400-atom (300-atom) model.  a) Total radial distribution functions. b) Angular distribution function resolved per central atomic species. The data are normalized to the number of triplets in each model. c) Distribution of coordination numbers resolved per chemical species, obtained by integrating the pair correlation functions up to a bonding cutoff of 3.0 \AA. d) Distribution of the local order parameter $q$ for tetrahedricity for four-coordinated Ge atoms.}
    \label{fig:Geamorphous}
\end{figure}

\begin{figure}[H]
 \centering
 \includegraphics[width=0.6\textwidth, keepaspectratio]{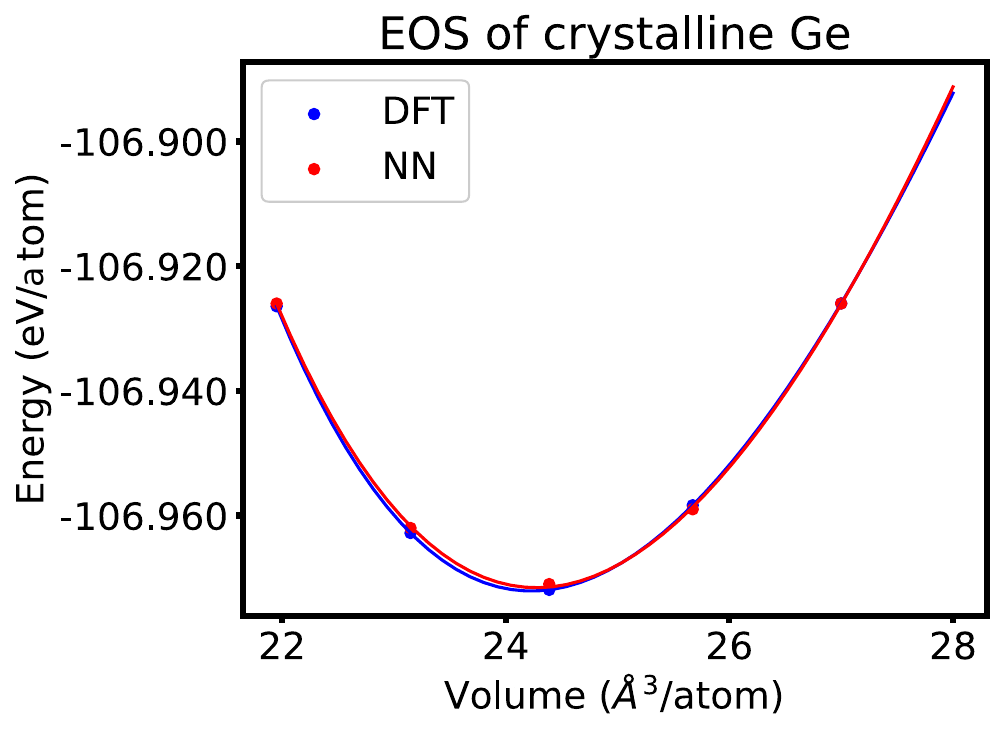}
 \caption{Equation of state (EOS) of the crystalline phase of Ge from DFT and NN
calculations.}
 \label{fig:GeEOSHex}
\end{figure}

The structural properties of  liquid Ge at the experimental density of the amorphous phase  (0.0438 atoms/\AA$^3$)\cite{baribeau} obtained from NN-MD simulation with a 2400-atom model at 1250 K are compared with those obtained from DFT-MD simulation with a 300-atom model from Ref. \cite{baratella2025} at the same density (see Fig. \ref{fig:Geliquid}).  The structural properties were averaged over a trajectory  50 ps long.  The pair correlation functions, the bond angles distribution function and the distribution of coordination numbers are compared with DFT results in Fig. \ref{fig:Geliquid}. Then we validated the NN potential on the structural properties of the amorphous phase at the same density. The 2400-atom liquid model was quenched from the melt at 1250 K to 300 K in 100 ps, and then equilibrated at 300 K for 40 ps.  Structural properties are averaged over a subsequent 50 ps long. In Fig. \ref{fig:Geamorphous}, the structural properties from NN-MD are compared with those obtained from an amorphous model generated by DFT-MD simulation in Ref. \cite{baratella2025} with a similar protocol. The pair correlation functions, the bond angles distribution function, the distribution of coordination numbers and the distribution of the local order parameter q for tetrahedricity for four-coordinated Ge atom are compared with DFT results in Fig. \ref{fig:Geamorphous}. We also quantified the fraction of four-coordinated Ge atoms in tetrahedral geometry by integrating the distribution of q by integrating the q parameter between 0.8 and 1 as discussed in previously, resulting in 64\% in NN model compared to 68\% in DFT results from Ref. \cite{baratella2025}. Finally, the NN potential was validated on the structural properties of the diamond crystal phase of Ge by fitting the EOS (see  Figs. \ref{fig:GeEOSHex}). The resulting parameters at equilibrium are compared with DFT data in Tables \ref{tab:GepEOShex}.

\begin{center}
\captionsetup{type=Table}
\captionof{table}{Fitting parameters of the Birch-Murnaghan equation of state of crystalline Ge  from NN and DFT-PBE calculations at zero temperature. The energy ($E_0$), volume ($V_0$), bulk modulus ($B$) and derivative of $B$ ($B^\prime$) at equilibrium are reported.}
\label{tab:GepEOShex}
\begin{tabular}{c c c c c}

\hline
 & $E_0$ (eV atom$^{-1}$)& $V_0$ (\AA$^3$ atom$^{-1}$)& $B$ (GPa) & $B^\prime$ \\
\hline

DFT & -106.972 &24.22 &57.58 &4.64\\
NN & -106.972 &24.27 &56.20 &3.60\\

\hline
\\
\end{tabular}

\end{center}{}

\subsection{$Sb$}
\begin{figure}[H]
    \centering
    \includegraphics[width=0.4\linewidth,keepaspectratio]{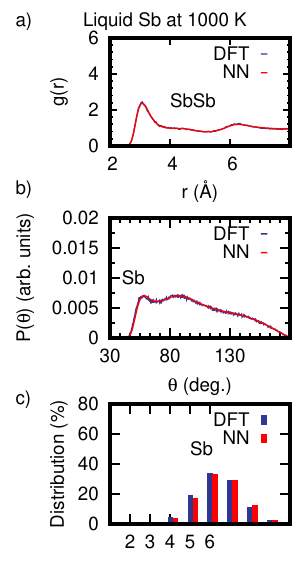}
    \caption{Structural properties of liquid Sb at 1000 K from NN (red curves) and DFT (blue curves) simulations. NN (DFT) data refer  to a 1000-atom (144-atom) model.   a) Total and partial radial distribution functions. b) Angular distribution function resolved per central atomic species. The data are normalized to the number of triplets in each model. c) Distribution of coordination numbers resolved per chemical species, obtained by integrating the pair correlation functions up to a bonding cutoff of 3.66 \AA. We used the same bonding cutoff of the amorphous phase.}
    \label{fig:Sbliquid}
\end{figure}

\begin{figure}[H]
    \centering
    \includegraphics[width=0.4\linewidth,keepaspectratio]{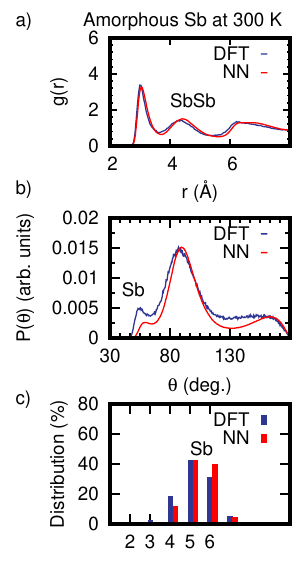}
    \caption{Structural properties of amorphous Sb at 300 K from NN (red curves) and DFT (blue curves) simulations. NN (DFT) data refer  to a 1000-atom (144-atom) model.  a) Total and partial radial distribution functions. b) Angular distribution function resolved per central atomic species. The data are normalized to the number of triplets in each model. c) Distribution of coordination numbers resolved per chemical species, obtained by integrating the pair correlation functions up to a bonding cutoff corresponding of 3.5 \AA.}
    \label{fig:Sbamorphous}
\end{figure}

\begin{figure}[H]
 \centering
 \includegraphics[width=0.6\textwidth, keepaspectratio]{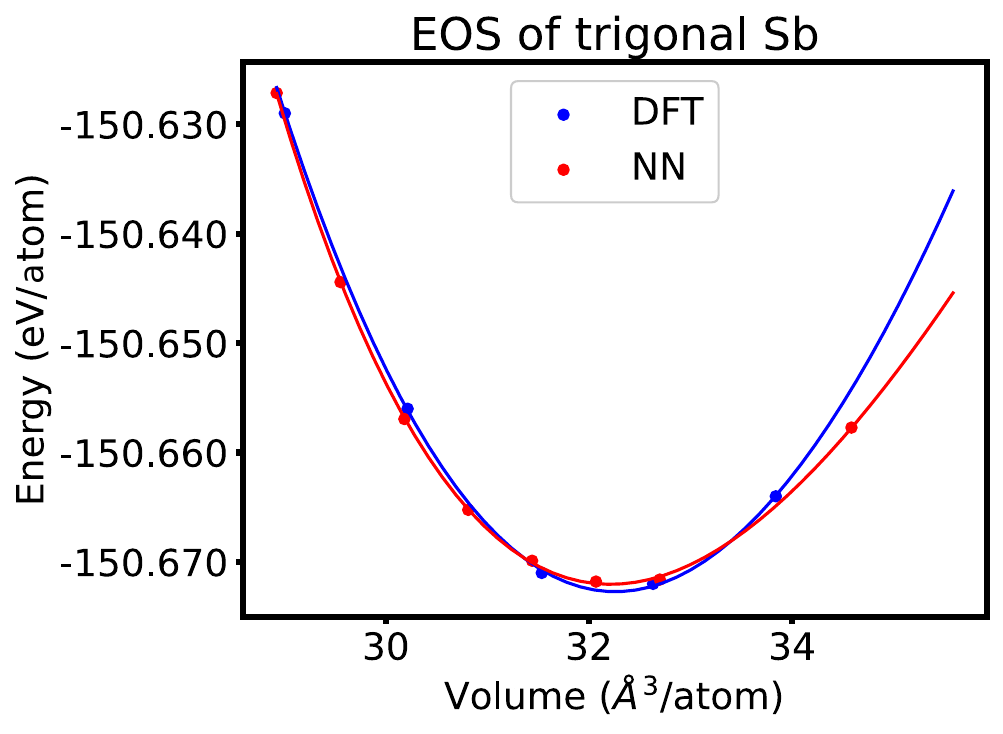}
 \caption{Equation of state (EOS) of the trigonal phase of Sb from DFT and NN
calculations.}
 \label{fig:SbEOSHex}
\end{figure}

The structural properties of  liquid Sb at the experimental density at melting  (0.0320 atoms/\AA$^3$)\cite{crawley1972density} obtained from NN-MD simulation with a 1000-atom model at 1000 K and compared with those obtained from DFT-MD simulation with a 144-atom model from Ref. \cite{dragoni2021mechanism} at the same density (see Fig. \ref{fig:Sbliquid}).  The structural properties were averaged over a trajectory  50 ps long.  The pair correlation functions, the bond angles distribution function and the distribution of coordination numbers are compared with DFT results in Fig. \ref{fig:Sbliquid}. Then, we validated the NN potential on the structural properties of the amorphous phase at the same density. The 1000-atom liquid model was quenched from the melt at 1000 K to 300 K in 100 ps, and equilibrated at 300 K for 40 ps.  Structural properties are averaged  over a subsequent run 50 ps long. In Fig. \ref{fig:Sbamorphous}, structural properties from NN-MD are compared with those obtained from an amorphous model generated by DFT-MD simulation in Ref. \cite{dragoni2021mechanism} with a similar protocol. The pair correlation functions, the bond angles distribution function, and the distribution of coordination numbers. Finally, the NN potential was validated on the structural properties of the trigonal crystal phase of Sb by fitting the EOS (see  Figs. \ref{fig:SbEOSHex}). The resulting parameters at equilibrium are compared with DFT data in Tables \ref{tab:SbpEOShex}.

\begin{center}
\captionsetup{type=Table}
\captionof{table}{Fitting parameters of the Birch-Murnaghan equation of state of hexagonal Sb  from NN and DFT-PBE calculations at zero temperature. The energy ($E_0$), volume ($V_0$), bulk modulus ($B$) and derivative of $B$ ($B^\prime$) at equilibrium are reported.}
\label{tab:SbpEOShex}
\begin{tabular}{c c c c c}

\hline
 & $E_0$ (eV atom$^{-1}$)& $V_0$ (\AA$^3$ atom$^{-1}$)& $B$ (GPa) & $B^\prime$ \\
\hline
DFT & -150.673 &32.26 &38.2 &2.3\\
NN  & -150.672 &32.21 &31.9 &7.3\\

\hline
\\
\end{tabular}

\end{center}{}

\subsection{$Te$}

\begin{figure}[H]
    \centering
    \includegraphics[width=0.4\linewidth,keepaspectratio]{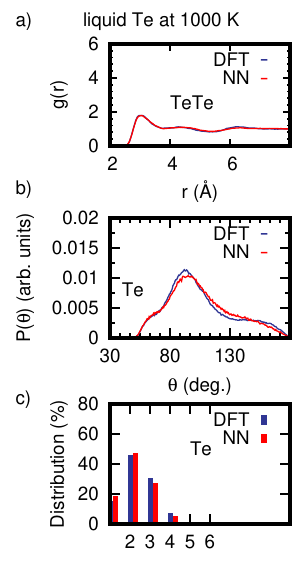}
    \caption{Structural properties of liquid Te at 1000 K from NN (red curves) and DFT (blue curves) simulations. NN (DFT) data refer  to a 1000-atom (200-atom) model. Total and partial radial distribution functions. b Angular distribution function resolved per central atomic species. The data are normalized to the number of triplets in each model. c) Distribution of coordination numbers resolved per chemical species, obtained by integrating the pair correlation functions up to a bonding cutoff of 3.2 \AA.}
    \label{fig:Teliquid}
\end{figure}

\begin{figure}[H]
 \centering
 \includegraphics[width=0.6\textwidth, keepaspectratio]{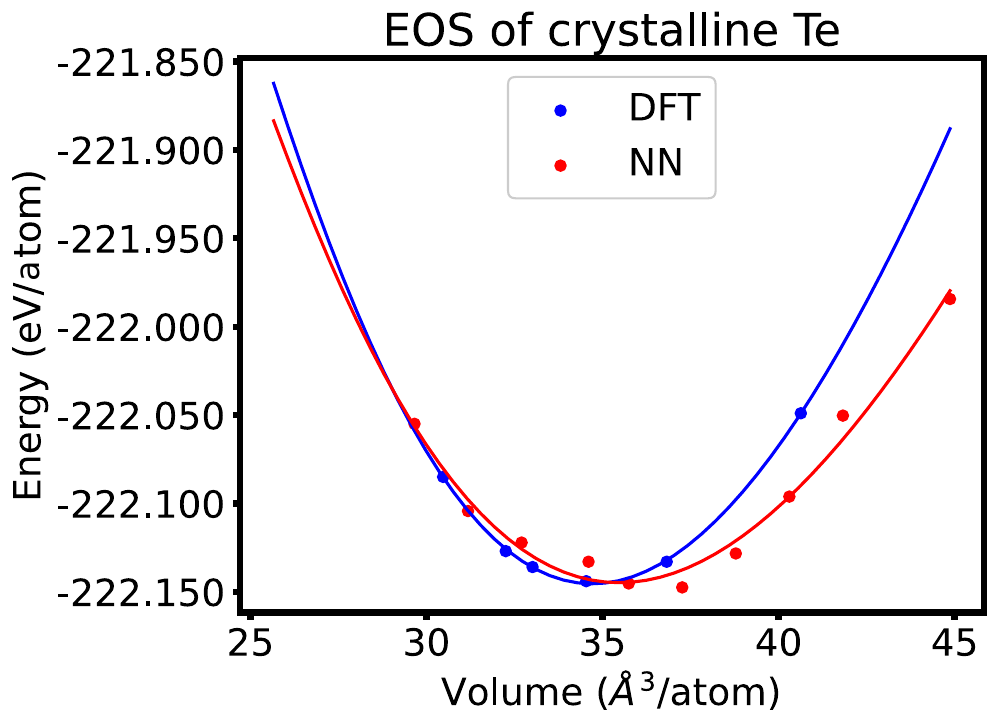}
 \caption{Equation of state (EOS) of the crystalline phase of Te from DFT and NN
calculations.}
 \label{fig:TeEOSHex}
\end{figure}

The structural properties of the liquid Te at the experimental density at melting  (0.0268 atoms/\AA$^3$) \cite{li2005thermophysical} obtained from NN-MD simulation with a 1000-atom model at 1000 K and compared with those obtained from DFT-MD simulation with a 200-atom model at the same density (see Fig. \ref{fig:Teliquid}).  The structural properties were averaged over a trajectory  50 ps long.  The pair correlation functions, the bond angles distribution function and the distribution of coordination numbers are compared with DFT results in Fig. \ref{fig:Teliquid}. Finally, the NN potential was validated on the structural properties of the crystalline phase of Te by fitting the EOS (see  Figs. \ref{fig:TeEOSHex}). The resulting parameters at equilibrium are compared with DFT data in Tables \ref{tab:TepEOShex}.

\begin{center}
\captionsetup{type=Table}
\captionof{table}{Fitting parameters of the Birch-Murnaghan equation of state of the crystalline phase of Te from NN and DFT-PBE calculations at zero temperature. The energy ($E_0$), volume ($V_0$), bulk modulus ($B$) and derivative of $B$ ($B^\prime$) at equilibrium are reported.}
\label{tab:TepEOShex}
\begin{tabular}{c c c c c}

\hline
 & $E_0$ (eV atom$^{-1}$)& $V_0$ (\AA$^3$ atom$^{-1}$)& $B$ (GPa) & $B^\prime$ \\
\hline
DFT & -222.145 &34.71 &34.6 &1.1\\
NN  & -222.145 &35.47 &26.5 &1.4\\

\hline
\\
\end{tabular}

\end{center}{}

\newpage
\section{Transferability}
Beside describing the structural properties of GeSbTe alloys included in the training set, we have tested the transferability of the MLIP to alloys outside the training set. Hereafter, we report the structural properties of the GST725 and Ge$_{9.{\bar 6}}$Sb$_2$Te$_5$ alloys whose crystallization kinetics was addressed by NN-MD simulations in the article, and two other alloys along the Ge-Sb$_2$Te$_3$ tie-line.

\subsection{$Ge_7Sb_2Te_5$}
\begin{figure}[H]
    \centering
    \includegraphics[width=0.8\linewidth,keepaspectratio]{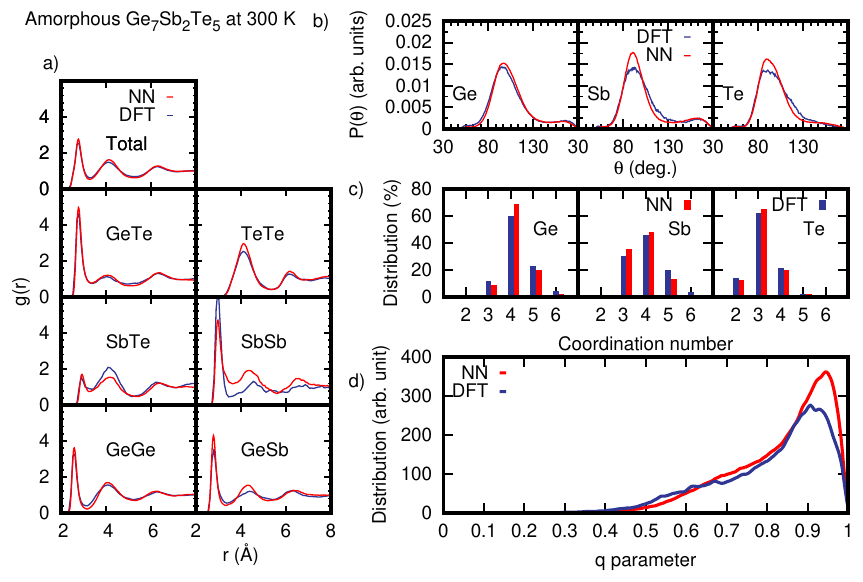}
    \caption{Structural properties of amorphous GST725 at 300 K from NN (red curves) and DFT (blue curves) simulations. NN (DFT) data refer  to a 1008-atom (210-atom) model. a) Total and partial radial distribution functions. b) Angular distribution function resolved per central atomic species. The data are normalized to the number of triplets in each model. c) Distribution of coordination numbers resolved per chemical species, obtained by integrating the pair correlation functions up to a bonding cutoff corresponding to 3.2 \AA\ for all pairs except for Sb-Te for which a longer cutoff  of 3.4\AA\ was used.  d) Distribution of the local order parameter $q$ for tetrahedricity for four-coordinated Ge atoms.}
    \label{fig:725amorphous}
\end{figure}

In the lack of experimental data for the density of $Ge_7Sb_2Te_5$,  we have modeled this alloy in a 1008-atom cubic cell at the theoretical density of  the liquid at 1000 K of 0.0335 atom/\AA$^3$, obtained from a DFT-NPT simulation with the NN potential. 
The liquid model was quenched from the melt at 1000 K to 300 K in 100 ps, and equilibrated at 300 K for 40 ps.  Structural properties are averaged  over a subsequent run 50 ps long.
The structural properties of the amorphous phase  are compared with those obtained from DFT-MD simulation with a 200-atom model in Fig. \ref{fig:725amorphous}, including the pair correlation functions, the bond angles distribution function, the distribution of coordination numbers and the distribution of the local order parameter $q$ for tetrahedricity for four-coordinated Ge atom. The average coordination number for different pairs of atoms are given in Table \ref{tab:am725NN}. The fraction of Ge atoms in tetrahedral geometry is 42\% in the NN model to be compared with the value of 45\% in the DFT model.

The fraction of Ge atoms in tetrahedral geometry is 46\% in the NN model to be compared with the value of 40\% in the DFT model.

\begin{center}
\captionsetup{type=Table}
\captionof{table}{Average coordination numbers for different pairs of atoms in amorphous GST725 at 300 K generated from  DFT (in parenthesis) and NN simulations. Details on the simulation cells and bonding cutoff are given in the caption of Fig. \ref{fig:725amorphous}.
}
\label{tab:am725NN}

\begin{tabular}{ c c c c c c  }
\hline

-       & Ge          &        Sb  & Te \\ \hline
Total   & 4.15 (4.20) &3.77 (3.95) &3.11 (3.12)\\
With Ge & 1.60 (1.68) &2.09 (2.00) &2.73 (2.72)\\
With Sb & 0.60 (0.57) &0.72 (0.97) &0.38 (0.39)\\
With Te & 1.95 (1.94) &0.95 (0.98) &0.00 (0.00)\\
\hline
\\
\end{tabular}
\end{center}

\subsection{Ge$_{9.{\bar 6}}$Sb$_2$Te$_5$}

In the lack of experimental data on Ge$_{9.{\bar 6}}$Sb$_2$Te$_5$, we have modeled this alloy in a 1000-atom cubic cell at the theoretical density of 0.0359 atom/\AA$^3$. This density was obtained from a NPT simulation with the NN potential in the liquid at 1000 K. The initial density for this simulation was linearly extrapolated from the density of amorphous Ge and GST725 (0.0382 atom/\AA$^3$). After an  equilibration MD simulation 30 ps long, we obtained the theoretical density of 0.0359 atom/\AA$^3$ as the average over 30 ps. This value is slightly lower than the initial estimate.  The liquid model was quenched from the melt at 1000 K to 300 K in 100 ps, and equilibrated at 300 K for 40 ps.  Structural properties are averaged  over a subsequent run 50 ps long.

\begin{figure}[H]
    \centering
    \includegraphics[width=0.8\linewidth,keepaspectratio]{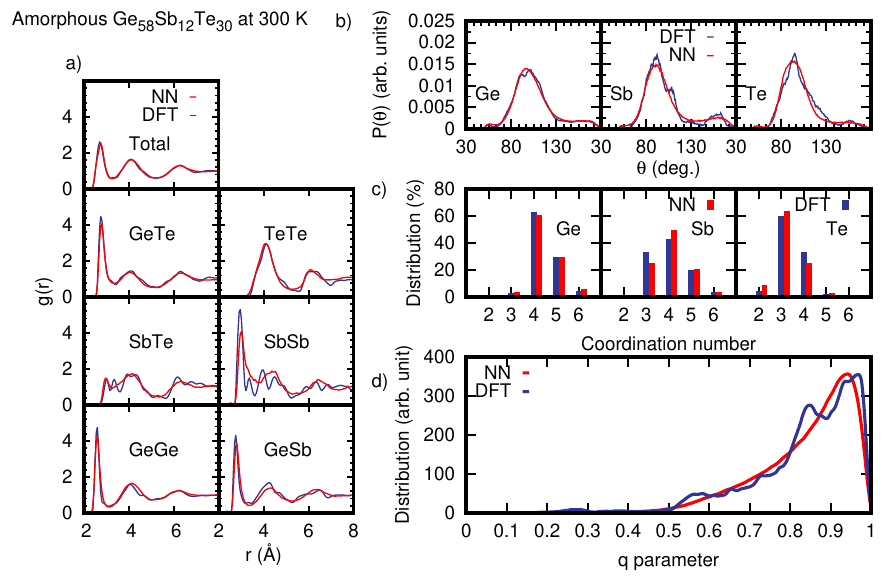}
    \caption{Structural properties of amorphous Ge$_{9.{\bar 6}}$Sb$_2$Te$_5$ (Ge$_{58}$Sb$_{12}$Te$_{30}$) at 300 K from NN (red curves) and DFT (blue curves) simulations.  NN (DFT) data refer  to a 1000-atom (210-atom) model. a) Total and partial radial distribution functions. b Angular distribution function resolved per central atomic species. The data are normalized to the number of triplets in each model. c Distribution of coordination numbers resolved per chemical species, obtained by integrating the pair correlation functions up to a bonding cutoff of 3.2 \AA\ for all pairs except for Sb-Te for which a longer cutoff  of 3.4 \AA\ was used. d) Distribution of the local order parameter $q$ for tetrahedricity for four-coordinated Ge atoms.}
    \label{fig:a581230}
\end{figure}

The structural properties of the amorphous phase  from NN-MD simulation are  compared with those obtained from DFT-MD simulation with a 200-atom model in Fig. \ref{fig:a581230}, including the pair correlation functions, the bond angles distribution function, the distribution of coordination numbers and the distribution of the local order parameter $q$ for tetrahedricity for four-coordinated Ge atom. The average coordination number for different pairs of atoms are given in Table \ref{tab:a581230}. The fraction of Ge atoms in tetrahedral geometry is 46\% in the NN model to be compared with the value of 39\% in the DFT model.

\begin{center}
\captionsetup{type=Table}
\captionof{table}{Average coordination numbers for different pairs of atoms in amorphous Ge$_{9.{\bar 6}}$Sb$_2$Te$_5$ at 300 K generated from  DFT (in parenthesis) and NN simulations. Details on the simulation cells and bonding cutoff are given in the caption of Fig. \ref{fig:a581230}.
}
\label{tab:a581230}

\begin{tabular}{ c c c c c c  }
\hline

-       & Ge          &        Sb  & Te \\ \hline
Total   & 4.38 (4.36) &4.01 (3.94) &3.21 (3.12)\\
With Ge & 2.39 (2.29) &2.49 (2.44) &2.85 (2.72)\\
With Sb & 0.52 (0.51) &0.63 (0.69) &0.35 (0.39)\\
With Te & 1.47 (1.56) &0.89 (0.81) &0.01 (0.00)\\
\hline
\\
\end{tabular}
\end{center}

\subsection{Alloys along the Ge-Sb$_2$Te$_3$ tie-line}
\begin{figure}[H]
    \centering
    \includegraphics[width=0.8\linewidth,keepaspectratio]{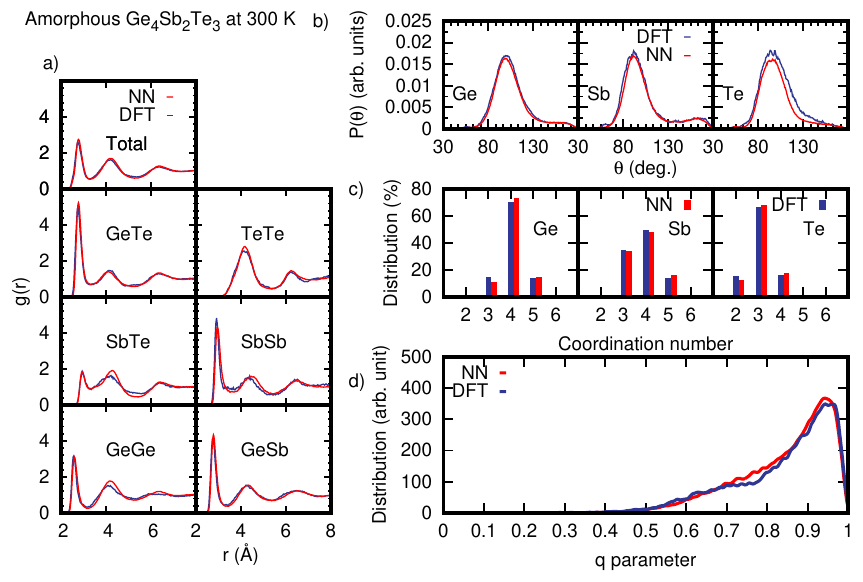}
    \caption{Structural properties of amorphous GST423 at 300 K from NN+D3 (red curves) and DFT+D3 (blue curves) simulations. NN (DFT) data refer  to a 864-atom (216-atom) model. a) Total and partial radial distribution functions. b) Angular distribution function resolved per central atomic species. The data are normalized to the number of triplets in each model. c) Distribution of coordination numbers resolved per chemical species, obtained by integrating the pair correlation functions up to a bonding cutoff of 3.2 \AA\ for all pairs except for Sb-Te for which a longer cutoff of 3.4 \AA\ was used. d) Distribution of the local order parameter $q$ for tetrahedricity for four-coordinated Ge atoms for the NN model.}
    \label{fig:423amorphousD3}
\end{figure}

\begin{figure}[H]
    \centering
    \includegraphics[width=0.8\linewidth,keepaspectratio]{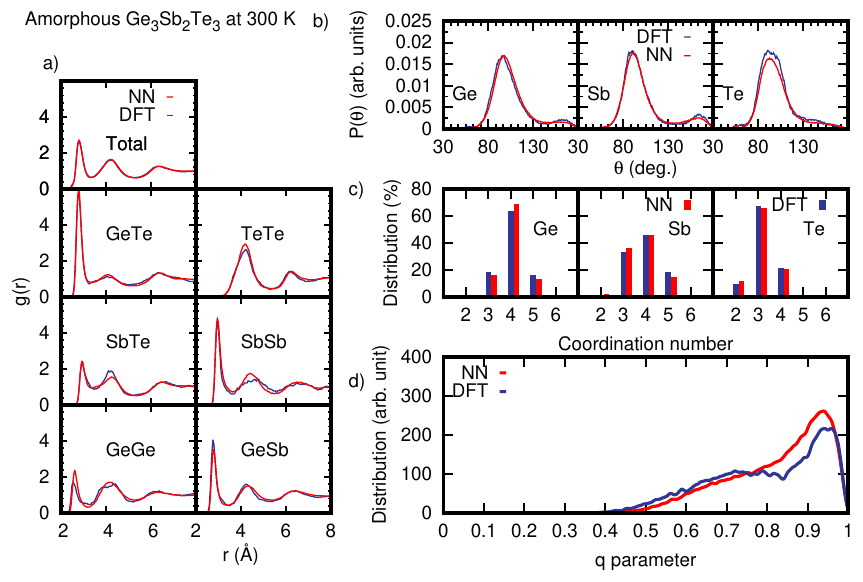}
    \caption{Structural properties of amorphous GST323 at 300 K from NN+D3 (red curves) and DFT+D3 (blue curves) simulations. NN (DFT) data refer  to a 864-atom (216-atom) model. a) Total and partial radial distribution functions. b) Angular distribution function resolved per central atomic species. The data are normalized to the number of triplets in each model. c) Distribution of coordination numbers resolved per chemical species, obtained by integrating the pair correlation functions up to a bonding cutoff of 3.2 \AA\ for all pairs except for Sb-Te for which a longer cutoff  of 3.4\AA\ was used.  d) Distribution of the local order parameter $q$ for tetrahedricity for four-coordinated Ge atoms.}
    \label{fig:323amorphousD3}
\end{figure}

As a final test on the transferability of the NN potential, in this section we report on the structural properties  of alloys along the Ge-Sb$_2$Te$_3$ tie-line which were investigated by DFT+D3 simulation in Ref. \cite{Omar2021}. Here, we compare the NN+D3 and DFT+D3 following the same protocol used in Ref. \cite{Omar2021} to generate amorphous models of  GST423 (see Fig. \ref{fig:423amorphousD3}) and GST323 (see Fig. \ref{fig:323amorphousD3}). 

\section{Crystallization of Ge-Sb-Te alloys}
\subsection{Crystallization of GST225}
\begin{figure}
    \centering
    \includegraphics[width=0.5\linewidth]{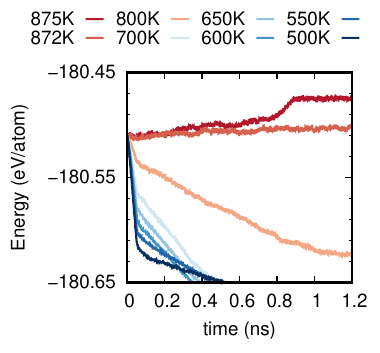}
    \caption{ The potential energy as a function of time at different temperatures of the cubic-liquid interface model of GST225. At 875 K the raise in the potential energy is due to the melting of the crystalline slab, while at 872 K the system reach a plateau suggesting to be very close to the melting temperature. }
    \label{fig:crys225}
\end{figure}

\begin{figure}
    \centering
    \includegraphics[width=0.45\linewidth]{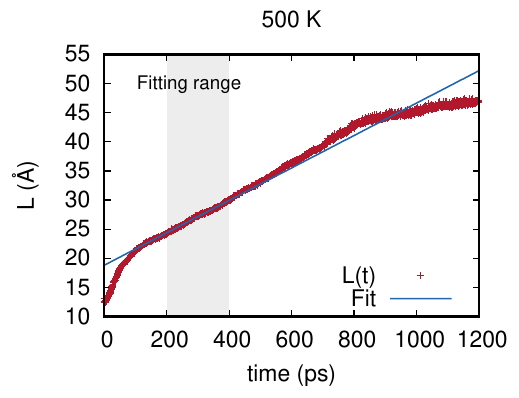}
    \includegraphics[width=0.45\linewidth]{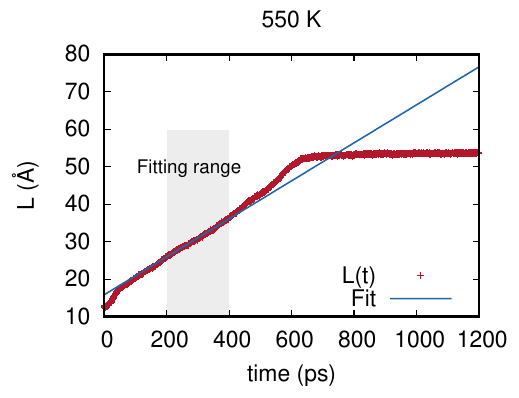}
    \includegraphics[width=0.45\linewidth]{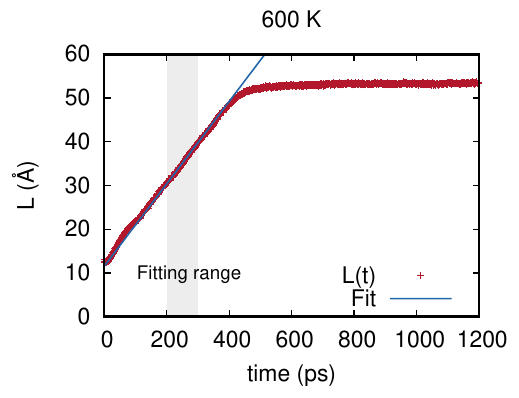}
    \includegraphics[width=0.45\linewidth]{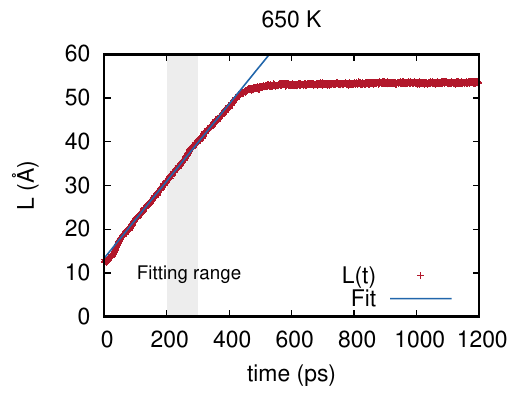}
    \includegraphics[width=0.45\linewidth]{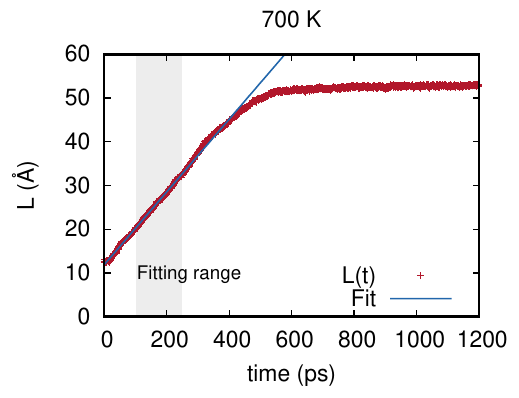}
    \includegraphics[width=0.45\linewidth]{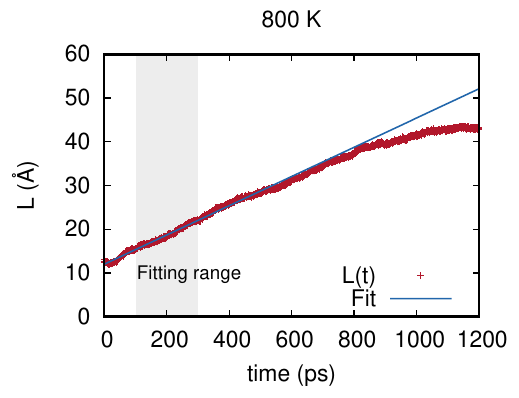}

    \caption{The evolution of the the effective (half) thickness of the crystalline slab $L(t)$ in simulations of GST225 at different temperatures and their linear fit in the range highlighted in gray which yields the crystal growth velocity.}
    \label{fig:crys225vg}
\end{figure}

Before investigating the crystallization with phase separation in GGST alloys, we first validated the NN potential on the crystallization of GST225 by computing the crystal growth velocity in the supercooled liquid. To this aim, we used 12960-atom model of the cubic-liquid interface from Ref. \cite{npjOmar} to simulate heterogeneous crystallization. The crystal-liquid interface in our model corresponds to the (001) surface of the cubic crystal. We performed several independent constant volume simulations in the range 500–875 K. 
The potential energy as a function of time is reported in Fig. \ref{fig:crys225}, from which we can se that the melting temperature is 872 K, close to the experimental value of 900 K \cite{mazzarelloliq}.  From these simulation we estimated the crystal growth velocity v$_g$ = $\frac{dL(t)}{dt}$, where $L(t)$ is the effective thickness of the crystalline region is given by $L(t)= N(t)/2A\rho$ . A is the cross section of the simulation cell and the factor two takes into account the presence of two growing surfaces. From the time evolution of L(t) we can extract the v$_g$ by as the slope of the linear region ( see Fig. \ref{fig:crys225vg}). The result on v$_g$ are given in Fig. 3 in the article.

\subsection{Crystallization with phase separation of GST725}
We simulated the crystallization and  phase separation of GST725 with a 8064-atom cell at the theoretical density of the liquid phase at 1000 K estimated form DFT-NPT simulation (0.0335 atom/\AA$^3$). By assuming that GST725 will decompose into the thermodynamically favored products (Ge and GST225), and that Ge will segregate into the amorphous phase at its equilibrium density, the density of the remaining GST225 of 0.0296 atom/\AA$^3$  is 4\% lower than the equilibrium density of amorphous GST225 (0.0309 atom/\AA$^3$). 
We started from a model equilibrated at 1400 K for 20 ps, followed by a second equilibration at 1000 K for 20 ps, and a rapid quenching to 600 K in 40 ps. Once we reached the target temperature,  we simulated the crystallization with phase separation for about 7 ns at NVT conditions. 
The cluster analysis (see Sec. II in the article) yielded the same 
environments identified in GST523 (see Table I in the article), but for interfacial Te atoms that were not found in this latter alloy.

\begin{figure}[H]
    \centering
    \includegraphics[width=0.65\linewidth,keepaspectratio]{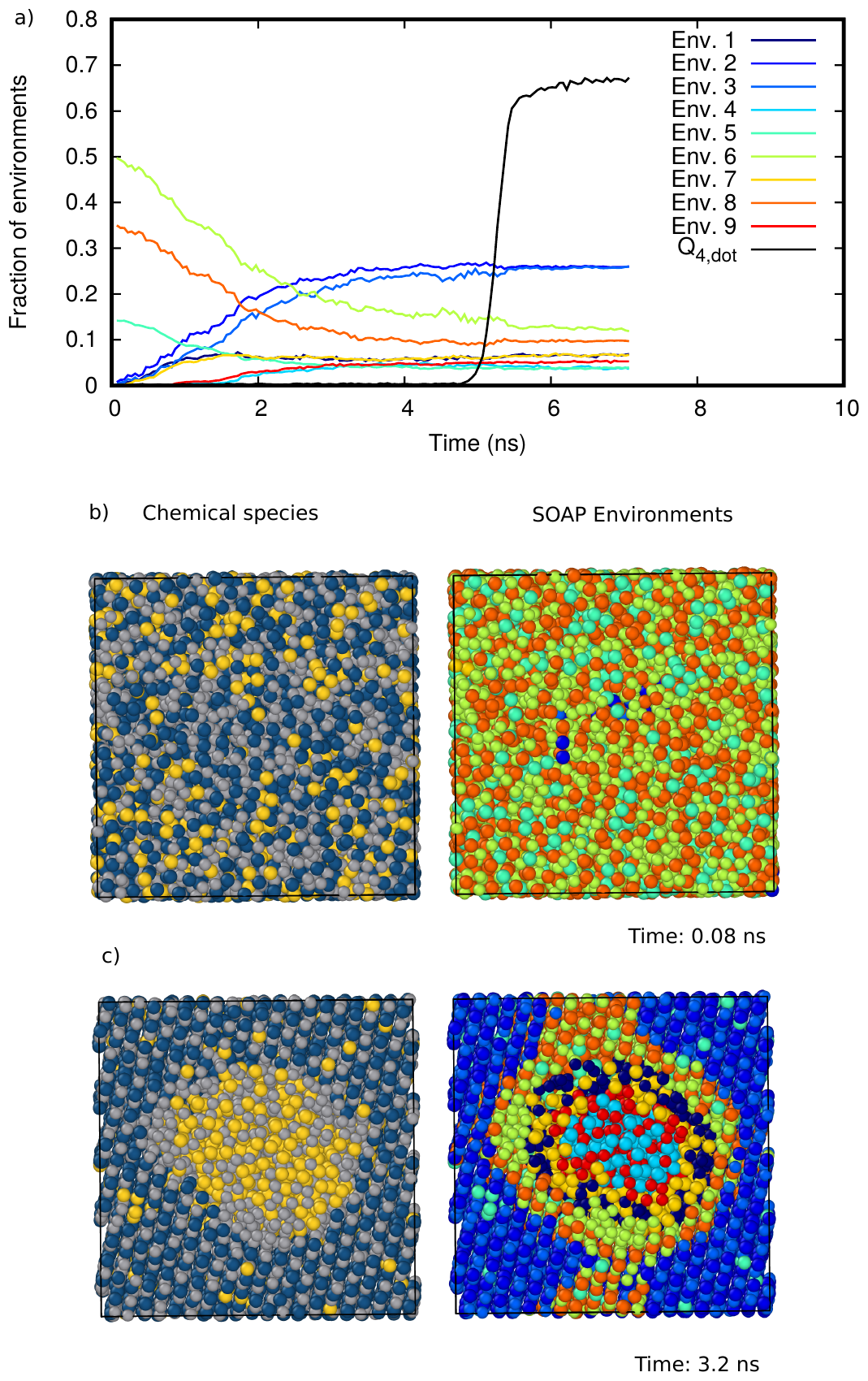}
     \caption{ Evolution in time of the fraction of atoms in the ten different SOAP environments during the crystallization of GST725. The description of the different environments is given in Table I in the article. Snapshot of the transformation process at b) the beginning and c) at the end of the simulation. In the panels on the left atoms are colored according to the chemical species (Ge, Sb, Te in gray, yellow and blue). In the panels on the right, atoms are colored according to the different SOAP environments whose color code is given in panel a).}
    \label{fig:crys725}
\end{figure}

The time evolution of the fraction of atoms belonging to the different ten SOAP environments is shown in Fig. \ref{fig:crys725}a together with the fraction of crystalline atoms identified by the Q$^{dot}_{4}$ order parameter for crystallinity. Snapshots at an early stage (0.08 ns), and at the end of the simulation are shown in Fig. \ref{fig:crys725}b-c which reports  the atoms colored  according to the chemical species on the left and according to the SOAP environment on the right.
The results are very similar to those of GST523 and GST9.625.
First, a separation into amorphous (supercooled liquid)  GeTe-like and Ge-Sb-like regions is observed and then, after the phase separation was completed, crystal nucleation and growth was observed within the GeTe-like regions. A slightly Sb-doped cubic GeTe crystal percolates through the simulation cell along the three directions (see Fig. \ref{fig:crys725}b-c).

The products of the crystallization and  phase separation of GST725 on the ns time scale  are thus crystalline Ge$_{12}$SbTe$_{12}$ (slightly Sb-doped GeTe) and amorphous Ge$_{54}$Sb$_{46}$, instead of the thermodynamically favored Ge and GST225 crystals.

\bibliographystyle{elsarticle-num} 
\bibliography{ref.bib}